\documentclass[preprint,authoryear,a4paper]{elsarticle}

\usepackage{verbatim}

\usepackage{geometry}

\usepackage{color,soul}

\usepackage[hidelinks]{hyperref}
\hypersetup{
  colorlinks   = true, 
  urlcolor     = blue,
  linkcolor    = blue, 
  citecolor    = blue 
}

\usepackage{graphicx}
\graphicspath{{./Figures/}}
\DeclareGraphicsExtensions{.pdf,.png,.jpg}
\usepackage{placeins}
\usepackage{caption}
\usepackage{subcaption}

\usepackage{booktabs}
\usepackage[table,xcdraw]{xcolor}
\usepackage{graphicx}
\usepackage{adjustbox}
\usepackage{multirow}

\usepackage{amssymb}
\usepackage{gensymb}
\usepackage[mathscr]{euscript}
\usepackage{bm}
\usepackage{mathtools}
\usepackage{nccmath}
\usepackage{amsmath}
\usepackage{braket}

\usepackage{algorithm}
\usepackage{algpseudocode}

\usepackage{lipsum}

\newcommand{\tabincell}[2]{\begin{tabular}{@{}#1@{}}#2\end{tabular}}
\journal{ISPRS Journal of Photogrammetry and Remote Sensing}

\begin{document}

\begin{frontmatter}

    \title{A Novel Full-Polarization SAR Images Ship Detector Based on the Scattering Mechanisms and the Wave Polarization Anisotropy}

    \author[swjtu]{Chuan Zhang}
    \author[swjtu]{Gui Gao}
    \author[swjtu]{Linlin Zhang}
    \author[nudt]{C. Chen}
    \author[hunan]{S. Gao\corref{cor1}}
    \author[NAU]{Libo Yao}
    \author[swjtu]{Qilin Bai}
    \author[swjtu]{Shiquan Gou}

    \cortext[cor1]{Corresponding Author: gaosh8899@126.com}
    \address[swjtu]{Faculty of Geosciences and Environmental Engineering, Southwest Jiaotong University, Chengdu, China}
    \address[nudt]{ the National University of Defense Technology, Changsha, China}
    \address[hunan]{ Hunan University, Changsha, China}
    \address[NAU]{Naval Aviation University of China}
    \begin{abstract}     
       Synthetic aperture radar (SAR) is considered being a good option for earth observation with its unique advantages. In this paper, we proposed an adaptive ship detector using full-polarization SAR images. First, by thoroughly investigating the scattering characteristics between ships and their background, and the wave polarization anisotropy, a novel ship detector is proposed by jointing the two characteristics, named Scattering-Anisotropy joint (joint-SA). Based on the theoretical analysis, we showed that the joint-SA is an effective physical quantity to show the difference between the ship and its background, and thus joint-SA can be used for ship detection of full-polarization image data. Second, the generalized Gamma distribution (G$\Gamma$D) was used to characterize the joint-SA statistics of sea clutter with a large range of homogeneity. As a result, an adaptive constant false alarm rate (CFAR) method was implemented based on the joint-SA. Finally, RADARSAT-2 and GF-3 data in C-band and ALOS data in L-band are used for verification. We tested on five datasets, and the experimental results verify the correctness and superiority of the constant false alarm rate (CFAR) method based on the joint-SA. In addition, the experimental results also showed that the signal-clutter ratio (SCR) of the proposed ship detector joint-SA (33.17 dB, 35.98 dB, 57.25 dB) is better than that of DBSP (8.92 dB, 3.43 dB, 25.40 dB) and RsDVH (17.28 dB, 11.17 dB, 54.55 dB). More importantly, the proposed detector joint-SA has higher detection accuracy and a lower false alarm rate.
    \end{abstract}

\end{frontmatter}


\section{Introduction}
\label{s:intro}

With the development of satellite technology, synthetic aperture radar (SAR) plays an increasingly important role in various fields, because of its capability in imaging day and night without considering the weather condition, such as land-cover classification \citep{ohki2018large}, change detection \citep{ciuonzo2017multiple}, disaster damage estimation \citep{chini2017hierarchical}, maritime traffic control \citep{renga2018segmentation}, inversion of surface parameters \citep{hajnsek2003inversion} and so on. Particularly, ship detection using SAR images is an effective means of ocean monitoring because the ship’s position information can help the government deal with illegal fishing \citep{snapir2019maritime}, illegal immigration, and so on. However, the detection of ships using SAR images is a complex and important topic, not only because of the sophistication of the ships themselves but also because of the sophistication and variability of the marine environment, especially for small targets, where the detection difficulty rises dramatically.  \par

In the past decades, algorithms of ship detection using SAR images data have been extensively researched, and it has got many credible results. Among the existing models, the constant false alarm rate (CFAR) detector based on the statistical model of the sea clutter is a classical ship detection model and is widely used \citep{robey1992cfar, wackerman2001automatic}. Meanwhile, Gaussian distributions, Weibull \citep{schleher1976radar}, Gamma \citep{principe1998target}, K \citep{jakeman1987generalized}, and Generalized K distribution \citep{ferrara2011generalized, liao2008using} were used to describe the statistical of the sea clutter. The generalized Gamma distribution \citep{martin2014statistical} (G$\Gamma$D) can get higher fitting accuracy than other statistical models and has been experimentally proven to be effective.\par
In the early stage of using single-polarization SAR image to detect, Yeremy et al. \citep{yeremy2001ocean} found that, at low incidence angles, the HV polarization has the highest SCR, which is the average value over the number of pixels, and the HH polarization has better SCR at high incidence angles. Tian et al. \citep{tian2015ship} proposed a ship detection method based on target enhancement and non-parametric clutter estimation using single-polarization SAR image, which not only improves the uniform image and non-uniform image of the ship-sea contrast, and adaptively estimated the clutter distribution in the enhanced image. However, single-polarization SAR data cannot provide rich polarization information, so its detection performance will be dramatically reduced in complex sea conditions.\par
Compared with the single-polarization SAR data, multi-polarization data contains richer information on backscattering and phase and provides a much more comprehensive characterization of the polarization information of complex sea conditions. Based on the dual-polarization SAR data, Gao et al. \citep{gao2013ship} proposed a ship CFAR detection method, which constructs a new PMA detector with improved ship SCR and makes it easier to detect ships from clutter. Shirvany et al.  \citep{shirvany2012ship} based on the fact that the degree of polarization (DoP) can describe the fundamental quantity of partially polarized electromagnetic fields, studied the performance of DoP  by combining ship and oil-spill detection under different polarization in hybrid, compact, and linear dual-pol SAR images.\par
Since the full-polarization data can completely reflect the scattering characteristics of the target, more and more researchers are interested in using full-polarization data for ship detection. Traditional ship detection algorithms, such as the total power detector (SPAN), the polarimetric whitening filter (PWF) \citep{liu2019cfar}, the power-maximization synthesis (PMS) \citep{chaney1990performance} detector, etc. directly fuse the intensity information of the scattering matrix to perform ship detection. Although these detectors can achieve excellent results in the sea state of calm, their detection effectiveness decreases sharply when the sea state is complex. Therefore, by taking full advantage of the rich information from the full-polarization SAR data, a detection method that combines the high-resolution capability of X-band sensors with the removal of azimuthal ambiguity for ship detection has been proposed by Velotto et al.\citep{velotto2013azimuth}. \par
However, even though these detectors have proved the ability to distinguish ships from sea clutter, the detection performance of intensity or amplitude-based methods is susceptible to sea conditions. Compared to the use of intensity of amplitude or intensity information, an alternative approach is to use polarization decomposition to explore new detectors. Using the similarity parameter (SP)\citep{yang2001similarity} of surface scattering, An et al.\citep{an2009similarity} added the term “1-SP” into the PMS to form the new detector. Marino et al.\citep{marino2012detecting} proposed a geometrical perturbation-polarimetric notch filter (GP-PNF) derived from the complex space of polarized targets to detect targets at sea. Yang et al.\citep{yang2019saliency} proposed PolSAR image saliency detection based on weighted perturbation filters. Zhang et al.\citep{zhang2019ship} proposed a new detection algorithm based on improving the geometrical GP-PNF. Based on non-negative matrix factorization (NMF), Guo et al.\citep{guo2011novel} proposed a new strategy for ship detection. This strategy uses non-negative eigenvalues to estimate sparsity from histograms that can reveal the sparse distribution of eigenvalues. Combining non-negative and sparse features, this method can be flexible and efficient for ship detection. Cloude et al.\citep{cloude1996review} used polarization parameters, such as entropy and eigenvalues, for ship detection. But it has been studied that these two detectors are not suitable for the detection of small targets in complex scenes. Zhang et al.\citep{zhang2017novel} used full-polarization SAR images to generate the polarimetric covariance difference matrix (CP) to improve the local contrast of the images and then extracted Shannon entropy features from them for the ship detection. The effectiveness of the method has been proved on the GF-3. \par
Ships and other man-made objects on the surface of the ocean are complex metallic structures with complex scattering mechanisms (SM). Freeman et al.\citep{freeman1998three} proposed a three-component model that decomposes the scattering power of the target as: surface scattering, due to direct backscattering from surfaces perpendicular to the radar beam; double-bounce scattering, because of the dihedral formed by the vertical ship’s conducting plates and the sea surface; volume scattering, formed because of multiple bouncing of electromagnetic waves \citep{velotto2013azimuth}. Based on the three-component model, Yamaguchi et al.\citep{yamaguchi2005four} proposed a four-component scattering model by adding the helix scattering characteristics. Through combining the double-bounce scattering, volume scattering, and helix scattering from the complete polarimetric covariance difference matrix, Zhang et al.\citep{zhang2018ship} proposed a new detector, named   (for simplicity, we call this detector “DBSP” in this paper). Experiments showed that the detection results outperformed the other ship detectors. Based on the DBSP detector, Zhang et al. \citep{zhang2020polsar} used the similarity parameter of the surface scattering to construct a new detector named RsDVH. The effectiveness of the detector RsDVH was proved experimentally. \par
Although the above detectors have experimentally proven their effectiveness, they still lose their effectiveness for some relatively weakly scattering targets, especially in the complex sea condition. More seriously, the robustness of exiting detectors is not good, and their detection accuracy is not guaranteed across different types of satellite platforms or different bands of data.\par
Overall, the existing ship detector can’t effectively detect ships, especially for some relatively weakly scattering targets on the condition of the high sea state. In this paper, by jointing the scattering characteristics of the target and the wave polarization anisotropy, we proposed a new full-polarization SAR image ship detector named joint-SA. Therefore, the contribution of this paper is:\par
1) From two different perspectives, a new ship detector joint-SA is proposed based on the target scattering characteristics and the wave polarization anisotropy. This detector can be seen as a joint of scattering power and wave polarization anisotropy to highlight the difference between the target and its background. From the perspective of SCR and detection accuracy, we also proved that the detector is an effective physical quantity for ship detection.\par
2) We tested and verified the suitability of the   for characterizing joint-SA statistics of sea clutter with a wide range of homogeneity. We also implemented adaptive detection of full-polarization SAR images by combining the CFAR algorithm and G$\Gamma$D.\par
3) The correctness of the theoretical analysis and the superiority of the CFAR method based on joint-SA detection were verified by experiments on the GF-3 in C-band, RADARSAT-2 in C-band, and ALOS in L-band.\par
The remainder of this paper is arranged as follows. Section \ref{theory} simply describes the theoretical background. In Section \ref{Method}, the details of detector joint-SA are stated carefully. Comparison experiments with different detectors are performed in Section \ref{s:Experiments}. Conclusions and discussion are given in Section \ref{s:Conclusion} .\par

\section{Theoretical Background}
\label{theory}

In this section, we mainly reviewed the basic theoretical knowledge in SAR image processing, including the scattering matrix [S], covariance matrix [C].

Full-Polarization (FP) SAR can measure the complete backscatter information of the target and comprehensively characterize the scattered power of the target, usually using a  scattering matrix \citep{guissard1994mueller} to characterize the information of each element of SAR image, which is defined as\citep{lee2017polarimetric}
\begin{equation}
	[S]=\left[\begin{matrix}
		S\textsubscript{HH} & S\textsubscript{HV} \\
		S\textsubscript{VH} & S\textsubscript{VV}
	\end{matrix}
	\right]
\end{equation}
where  $S\textsubscript{pq}$  denotes the p-transmitted and q-received scattering element. $H$ indicates horizontal polarization wave, $V$ indicates vertical polarization wave.\par
In real scenarios, the targets observed by SAR satellites are not ideal scattering mechanisms, but a combination of multiple objects called partial targets \citep{deschamps1973poincare}. The two-dimensional scattering matrix cannot effectively characterize all the properties of the target, so the second-order statistics of the scattering matrix are introduced \citep{marino2013notch}. We assume that the scattering matrix satisfies the reciprocity theorem. When k is expressed in the Pauli basis, i.e., [k]=[S\textsubscript{HH}, $\sqrt{2}$S\textsubscript{HV}, S\textsubscript{vv}] , then the covariance matrix can be expressed as 
\begin{equation}
	\begin{split}
		[C]=&\left \langle k \cdot k\textsuperscript{T} \right \rangle \\
	=&\left[\begin{matrix}
		\left \langle \left | S_{HH} \right |\textsuperscript{2} \right \rangle & \left \langle \sqrt{2}S_{HH} S_{HV}^{*}\right \rangle &  \left \langle S\textsubscript{HH}S_{VV}^{*}\right \rangle \\
		\left \langle \sqrt{2} S_{HV} S_{HH}\textsuperscript{*} \right \rangle & \left \langle 2 \left |S\textsubscript{HV} \right |\textsuperscript{2} \right \rangle & \left \langle \sqrt{2}S_{HV}  S_{VV}\textsuperscript{*}\right \rangle \\	  
		\left \langle S\textsubscript{VV}S_{HH}^{*}\right \rangle & \left \langle \sqrt{2}S\textsubscript{VV}S_{HH}^{*}\right \rangle & \left \langle \left | S\textsubscript{VV} \right |\textsuperscript{2} \right \rangle
	\end{matrix}
	\right] \\
	=&\left[\begin{matrix}
		C_{11} & C_{12} & C_{13}\\
		C_{21}^{*} & C_{22} & C_{23} \\
		C_{31}^{*} & C_{32}^{*} & C_{33}
	\end{matrix}
	\right]
	\end{split}
\end{equation}
where  $\left \langle \cdot \right \rangle$ represents ensemble averaging; superscript “*” is the complex conjugate; and  denotes the modulus of the complex signal.

\section{Scattering-Anisotropy Joint and Adaptive Detection}
\label{Method}
\subsection{Four-component Deconmpsition}
\label{s:decomponent}
To better characterize the scattering mechanism of the target, Yamaguchi et al.\citep{yamaguchi2005four} decomposed the scattering mechanism of the target into four components, namely surface scattering  $P\textsubscript{s}$, double-bounce scattering $P\textsubscript{d}$, volume scattering  $P\textsubscript{v}$, and helix scattering  $P\textsubscript{h}$. The decomposition of the equation is given by
\begin{equation}
    \begin{split}
		<[C]> = &f_{s}<[C]>_{s} +  f_{d}<[C]>_{d} +  f_{v}<[C]>_{v} +  f_{h}<[C]>_{h} \\
		=&f_{s} \left[\begin{matrix}
			|\beta|^{2} & 0 & \beta \\
			0           & 0 & 0 \\
			\beta^{*} & 0 & 1
		\end{matrix}\right] + f_{d} \left[\begin{matrix}
			|\alpha|^{2} & 0 & \alpha \\
			0           & 0 & 0 \\
			\alpha^{*} & 0 & 1
		\end{matrix}\right]   \\ &+ f_{v} \left[\begin{matrix}
			a & 0 & b \\
			0           & c & 0 \\
			d & 0 & e
		\end{matrix}\right] + \frac{f_{h}}{4}\left[\begin{matrix}
			1 & \pm j\sqrt{2} & -1 \\
			\mp j\sqrt{2}          & 2 & \pm j\sqrt{2} \\
			-1 & \mp j\sqrt{2} & 1
		\end{matrix}\right]
    \end{split}
\end{equation}
where the parameters a, b, c, d, and e are determined by 10$\log\frac{<|S_{VV}|^{2}>}{<|S_{HH}|^2>}$. In addition, the $P_{s}=f_{s}(1+|\beta|^{2})$, $P_{d}=f_{d}(1+|\alpha|^{2})$, $P_{v}=f_{v}$ and $P_{h}=f_{h}$; $P_{s}$, $P_{d}$, $P_{v}$, and $P_{h}$ should satisfy  $P_{t}=P_{s}+P_{d}+P_{v}+P_{h}=<|S_{HH}|^{2}+2|S_{HV}|^2+|S_{VV}|^2>$.\par
Inspired by\citep{yamaguchi2005four}, through utilizing the complete polarimetric covariance difference matrix [CP], Zhang et al.\citep{zhang2018ship} proposed a new detector $DBSP_{CP}$  (for convenience, we will call it “DBSP” in this paper)
\begin{equation}
	DBSP_{CP}  = ((P_d)_{CP} + (P_v)_{CP}) \cdot (P_h)_{CP}
\end{equation}
Although the effectiveness of the DBSP detector was proved on several data sets, the detector still has some drawbacks, as stated in \citep{zhang2020polsar}. First, the algorithm of DBSP removes the surface scattering feature, so the detector loses the ability to detect ships when they primarily exhibit surface scattering. Second, the [CP] is calculated based on the fact that the target pixel differs from its surrounding pixels in terms of the polarized scattering. That is, when the scattering difference between the target pixel and the surrounding pixels is not obvious, the value of the target pixels may be small or even zero, and then DBSP losses its detection effectiveness.\par
To overcome the disadvantages of the DBSP detector, Zhang et al.\citep{zhang2020polsar} proposed a new detector RsDVH 
\begin{equation}
	RsDVH=Rs_{max} \cdot ((P_d)_{max} + (P_v)_{max}) \cdot (P_h)_{max}
\end{equation}
where the subscript “max” indicates the selection of the maximum value from the values calculated by [C] and [CP], and $Rs_{max}$ ($0 \leq Rs_{max} \leq 1$) denotes the parameter associated with surface scattering. \par
Although the RsDVH detector is experimentally proven to be superior to the DBSP detector, our careful analysis of RsDVH reveals that there is a contradiction in equation (5). On the one hand, when the target pixel exhibits mainly surface scattering, that is to say, $Rs_{max} \rightarrow 1$, and the values of $(P_d)_{max}$, $(P_v)_{max}$, and $(P_h)_{max}$ are small. On the contrary,  when other scattering mechanisms prevail, the $Rs_{max} \rightarrow 0$, and the values of $(P_d)_{max}$, $(P_v)_{max}$, and $(P_h)_{max}$ are large. In other words, $Rs_{max}$ is inversely related to $((P_d)_{max} + (P_v)_{max}) * (P_h)_{max}$. On the other hand, when  $Rs_{max}$ and $((P_d)_{max} + (P_v)_{max}) * (P_h)_{max}$ are multiplied, the two do not serve to highlight the target pixels and suppress clutter pixels. This is the paradox of the detector RsDVH.

\subsection{Scattering-Anisotropy Joint}
\label{s:Joint-SA}
As described in section \ref{s:decomponent}, the ship detectors DBSP and RsDVH exist some disadvantages. To improve the detection of the accuracy of relatively weakly scattering targets, we urgently need a new characteristic independent of the target scattering mechanism which is joint with the target scattering characteristics to overcome the shortcomings in the detectors DBSP and RsDVH. Hence, we introduced a new physical quantity—wave polarization anisotropy $\Delta$S . This is for two reasons. First, the article\citep{zhang2018ship, zhang2020polsar} demonstrated that it is feasible to use the difference in scattering mechanisms between the ship and the clutter for detection. Second, the wave polarization anisotropy can characterize target non-stationary \citep{touzi2004ship}, which can compensate for the limitation of the detector DBSP and RsDVH depending only on the scattering characteristics. It should be noted that at incidence angles less than 60°, the values of $\Delta$S for the ocean are generally lower than those for ships, just as the paper \citep{touzi2004ship} stated.\par
As we all know, ships and other man-made objects at sea exhibit complex scattering mechanisms, and it is the sophistication of the target scattering mechanisms that can be described by the extreme value of the degree of polarization ($DoP$). Using the  $DoP$, Bicout et al.\citep{bicout1992multiply} introduced a measure of the wave entropy $S$ , which was given by
\begin{equation}
	S(DoP) = -ln(s(DoP))
\end{equation}
where $s(x)=\frac{1}{2}(1+x)^{\frac{1+x}{2}}(1-x)^{\frac{1-x}{2}}$ and $S(DoP)$ is is bijective, strictly decreasing function of $DoP$, and should satisfy $S(DoP=1) \leq S \leq S(DoP)$. The normalized value $S_n(DoP) = S(DoP)/S(0)$ takes values between 0 and 1. The variations of scattered wave purity $DoP$ with transmitting wave polarization produce a change of the wave entropy \citep{touzi2004ship}.\par
The $S(DoP)$ is directly related to $DoP$ which is characterizing the purity of the scattered wave. The dynamic range $\Delta$S is directly related to the dynamic range of $\Delta_{DoP}$ and thus can provide a measure of the complexity of the scattering mechanism \citep{touzi1992polarimetric}. This leads the expression for the wave polarization anisotropy was given \citep{touzi2000calibrated}
\begin{equation}
	\Delta S_n = S_n(DoP_{min}) - S_n(DoP_{max})
\end{equation}
where the extreme values of the   were derived by the method introduced by the Kostinski et al.\citep{kostinski1988optimal}. Therefore, we introduced wave polarization anisotropy $\Delta$$S_n$ to characterize the non-stationarity of the target. The larger the value of  $\Delta$$S_n$, the greater the variation of the signal with transmit-receive polarization \citep{touzi2004ship}.\par
As mentioned earlier, the scattering mechanisms alone cannot provide effective detection of relatively weakly scattering targets, such as detectors DBSP and RsDVH. Therefore, we proposed a new full-polarization SAR image ship detector joint-SA; and the expression of the joint-SA is given by
\begin{equation}
	joint-SA=(P_d + P_v) \cdot P_h \cdot \Delta S
\end{equation}
where $P_d$, $P_v$, $P_h$ denote double-bounce scattering, volume scattering, and helix scattering, respectively; and  $\Delta$$S_n$ represents the wave polarization anisotropy. It should be noted that  $P_d$, $P_v$, $P_h$ are calculated from [C] and not [CP]. There are two advantages to doing this. On the one hand, it can overcome the defects in the DBSP. On the other hand, it avoids complex operations and improves the efficiency of the algorithm.\par
Next, we will focus on explaining the rationality of the detector proposed in this paper:\par
1) From the point of view of the decomposition of scattering mechanism: as for the ocean, the most scattering mechanism is surface scattering; as for the target, because of the complex composition, the most scattering mechanism is the combination of the double-bounce scattering, volume scatter and helix scattering by the combination of two or more coherent scatters. Also, we can explain from equation (3) that the helix scattering power $P_h = 2 \cdot Im\left\{C_{12} + C_{23}\right\}$  is not present at the ocean surface because of the reflection symmetry of surface scattering, where both $C_{12}$ and $C_{23}$ are zero in $<[C]>_s$, while the corresponding in $<[C]>_h$  is complex \citep{zhang2018ship}. As a result, we can use helix scattering power as a reinforcement factor and do a multiplication operation directly with the sum of the double scattering and volume scattering powers to highlight the target-dominated scattering mechanism and suppress the background. \par
2) From the perspective of the wave polarization anisotropy: on the one hand, Touzi et al.\citep{touzi2004ship} have demonstrated experimentally that the wave polarization anisotropy can characterize the non-stationarity of the target; the larger the value of $\Delta$$S_n$, the greater the variation of the signal with transmit-receive polarization. On the other hand, when ships or other man-made objects are irradiated by electromagnetic waves, the variation of the signal with transmit-receive polarization is more significant, due to their complex structures and materials; however, the sea surface is relatively smooth, the change of signal with transmit-receive polarization is smaller. In other words, when  $\Delta$$S_n$ characterizes the target, i.e., $\Delta$$S_n \rightarrow 1$; and when $\Delta$$S_n$ characterizes the sea surface, i.e.,  $\Delta$$S_n \rightarrow 0$. Therefore, it is possible to characterize the difference between the target and the clutter from the perspective of the variation of the signal with the transmit-receive polarization. \par
3) From the perspective of combining the scattering and anisotropy characteristics: regardless of the scattering mechanism exhibited by the ship, at incidence angles below 60°, the wave polarization anisotropy of the sea surface is usually lower than that of a ship \citep{touzi2004ship}, that is, $(\Delta S_{n})_{ship} \geq (\Delta S_{n})_{sea}$. It should be noted that $0 < \Delta S_{n} < 1$, so we can consider $\Delta S_{n}$ as a weighting term that tends to 1 when $\Delta S_{n}$ characterizes the target, and tends to 0 when $\Delta S_{n}$ characterizes the clutter. In more detail, when the $\Delta S_{n}$ tends to 1, the $(P_d + P_v)*P_h *\Delta S$ tends to $(P_d + P_v)*P_h$, but when the$\Delta S_{n}$ tends to 0, the $(P_d + P_v)*P_h *\Delta S$ tends to 0. As a result, the combination of the two can effectively highlight the target and suppress clutter.\par
In summary, the detector proposed in this paper combines the scattering characteristics of the target with wave polarization anisotropy. Through our analysis, we found that even though the target exhibits mainly surface scattering, the combination of two features also can enhance the difference between the target and the sea clutter, helping us distinguish the target from the sea clutter. Moreover, the SCR is also used to evaluate the performance of the detector, and the expression is given by \citep{gao2018adaptive}
\begin{equation}
	SCR_{joint-SA}=20\log\frac{joint-SA_{T}}{joint-SA_{C}}
\end{equation}
where the subscripts T and C indicate the target and clutter pixels, respectively.
\begin{algorithm}[!htb]
	\label{algo:Detector}
	\caption{Adaptive CFAR Ship Detector Based on joint-SA}
	\begin{algorithmic}[1]
		\Require
		The original full-polarization SAR  data I; The temporary image T; The false alarm ratio $P_{fa}$, $\nu$, $\kappa$, $\sigma$, $i$, $j$;
		\Ensure
		The detection result R;
		\State Obtain the data size I;
		\State $[Row, Col] = siez(I)$;
		\State $T = zeros(Row, Col); R = zeros(Row, Col)$;
		\State Get the polarimetric covariance matrix [C] and 2$\times$2 scattering matrix [S];
		\For{each pixel $\in I$}
		\State Carry out the four-component decomposition by using polarimetric covariance matrix [C] and obtain $P_d$, $P_v$ and $P_h$;
		\State Use the 2x2 scattering matrix [S] getting the extreme of $DoP$, and then getting the wave polarization anisotropy $\Delta S$, i.e.:\\
		\quad \quad\quad \quad$\Delta S$ = $S_n(DoP_{min}) - S_n(DoP_{max})$
		\State Compute the joint-SA value: \\
		\quad \quad\quad \quad$joint-SA = (P_d + P_v)*P_h * \Delta S $
		\State Add the joint-SA value into T:  \\
		\quad\quad\quad\quad $T(i,j) = joint-SA$
		\EndFor
		\State Estimate the G$\Gamma$D parameters $v$, $\kappa$, $\sigma$;
		\State Set $P_{fa}$ and then compute the threshold m;
		\If{$\nu$ > 0}
		\State $m=\sigma(\frac{1}{\kappa}\Gamma^{-1}(1-P_{fa}))^{\frac{1}{v}}$
		\Else 
		\State $m=\sigma(\frac{1}{\kappa}\Gamma^{-1}(P_{fa},\kappa))^{\frac{1}{\nu}}$
		\EndIf
		\For{each pixel $q \in T$}
		\If{$T(q) > m$}
		\State Consider q as a ship pixel and then set the corresponding position of q in R to one, i.e.:
		\State $R(q) = 1$;
		\Else
		\State $R(q) = 0$;
		\EndIf
		\EndFor \\
		\Return the detection result R.
	\end{algorithmic}
\end{algorithm}

\subsection{G$\Gamma$D}
\label{s:GTD}
After the new detector was defined, the statistics of the detector need to be characterized to achieve adaptive detection 
of CFAR in different scenarios. In this paper, the G$\Gamma$D is used to characterize the statistical behavior of joint-SA. The reason for using   is that it has been widely used in different fields and has been proven to be an effective method for describing the statistical behavior of sea clutter \citep{ao2018detection, li2010efficient}. Li et al.\citep{li2011empirical} proposed a version of G$\Gamma$D with three parameters and the probability density function (pdf) is given:
\begin{equation}
	p(x) = \frac{|\nu|\kappa^\kappa}{\sigma\Gamma(\kappa)}(\frac{x}{\sigma})^{\kappa\nu-1},\sigma,|\nu|,\kappa,x>0
\end{equation}
where $\nu$, $\kappa$, $\sigma$ represent the power, shape, and scale parameters, and $\Gamma(\cdot)$ is the Gamma function.\par
Although the  G$\Gamma$D is an empirical model, it has also been shown to be relatively general \citep{li2011empirical}. Because the various distributions used to model SAR images can be considered as a special case, including Rayleigh  ($\nu$=2,$\kappa$=1 ), exponential ($\nu$=1,$\kappa$=2), Weibull ($\nu$=-2), and so on. So using the parameter estimation method of G$\Gamma$D in \citep{qin2012cfar}, the detection threshold can be expressed as:
\begin{equation}
	m=\left\{
	\begin{array}{lr}
		\sigma(\frac{1}{\kappa}\Gamma^{-1}(1-P_{fa}))^{\frac{1}{v}}, \nu\geq0 & \\
		\sigma(\frac{1}{\kappa}\Gamma^{-1}(P_{fa},\kappa))^{\frac{1}{\nu}}, \nu<0
	\end{array}
	\right.
\end{equation}
where $\Gamma(\cdot,\cdot)$  denotes the inverse incomplete Gamma function and  $P_{fa}$ is a given value of the false-alarm probability.

\section{Experimental Results and Analysis}
\label{s:Experiments}
In this section, experiments were performed on actual full-polarization SAR data sets which include C-band and L-band. Our goal is to certify the correctness of the theoretical analysis in Section \ref{Method}. First, the fitting abilities of the G$\Gamma$D for different detectors used for CFAR ship detection were accessed. Subsequently, we accessed if the detector based on joint-SA is more effective and can improve the SCR compared with other detectors. Finally, the CFAR detection performance of several detectors was compared. 
\subsection{Test Data}
\begin{figure}[H]
	\centering
	\subcaptionbox{Scene 1}[0.32\linewidth]{\includegraphics[width=\linewidth]{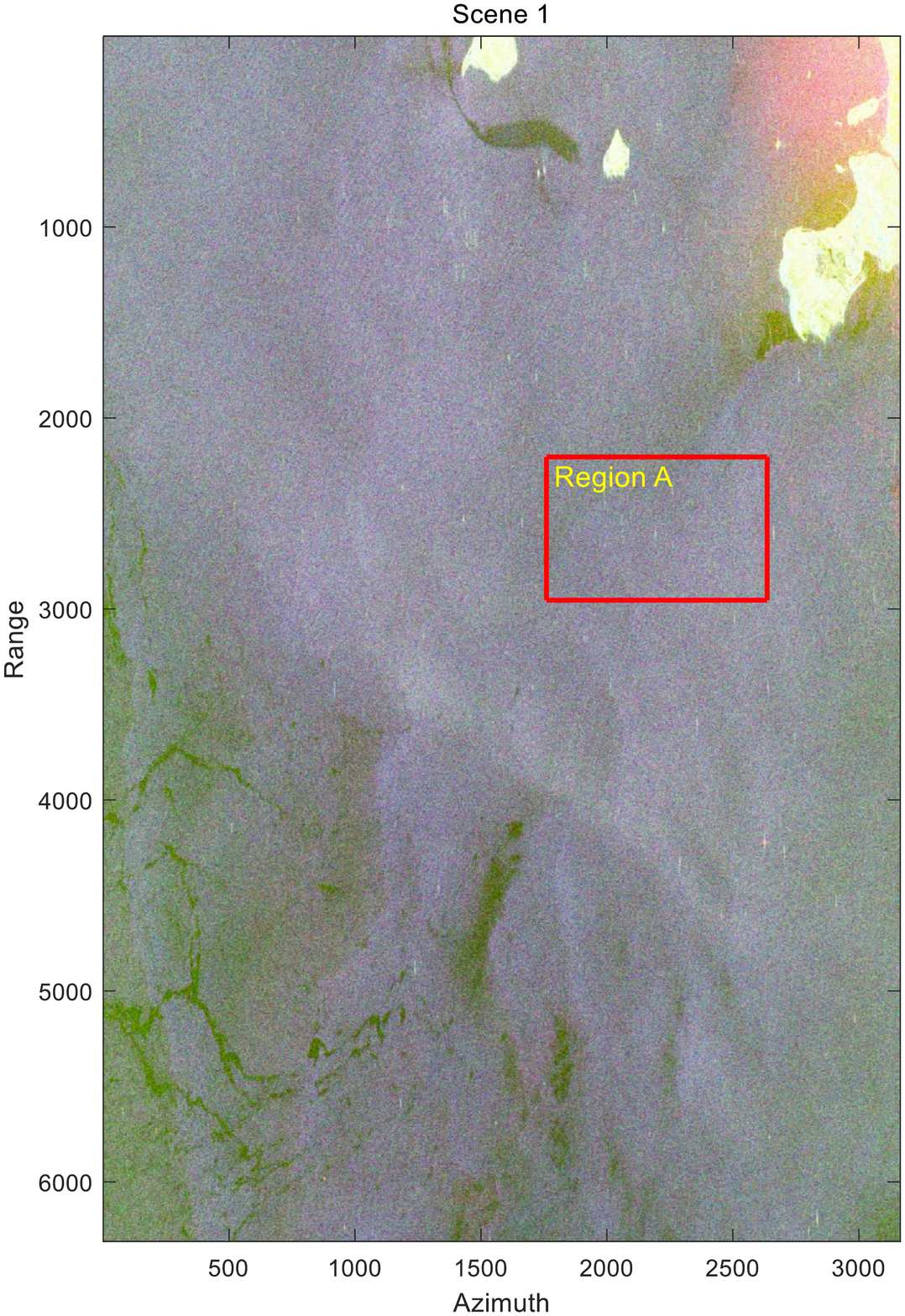}}
	\subcaptionbox{Scene 2}[0.32\linewidth]{\includegraphics[width=\linewidth]{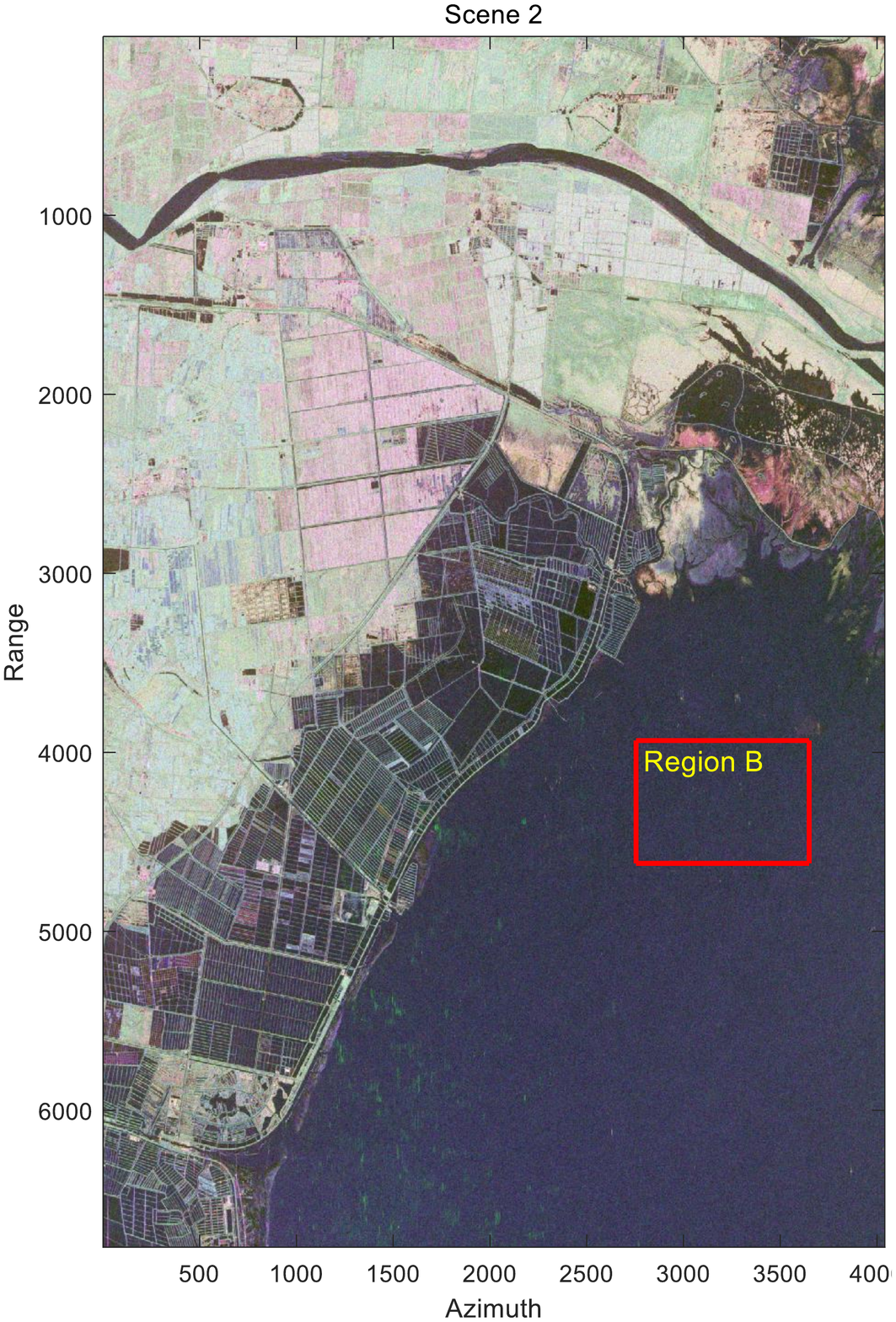}}
	\subcaptionbox{Scene 3}[0.32\linewidth]{\includegraphics[width=\linewidth]{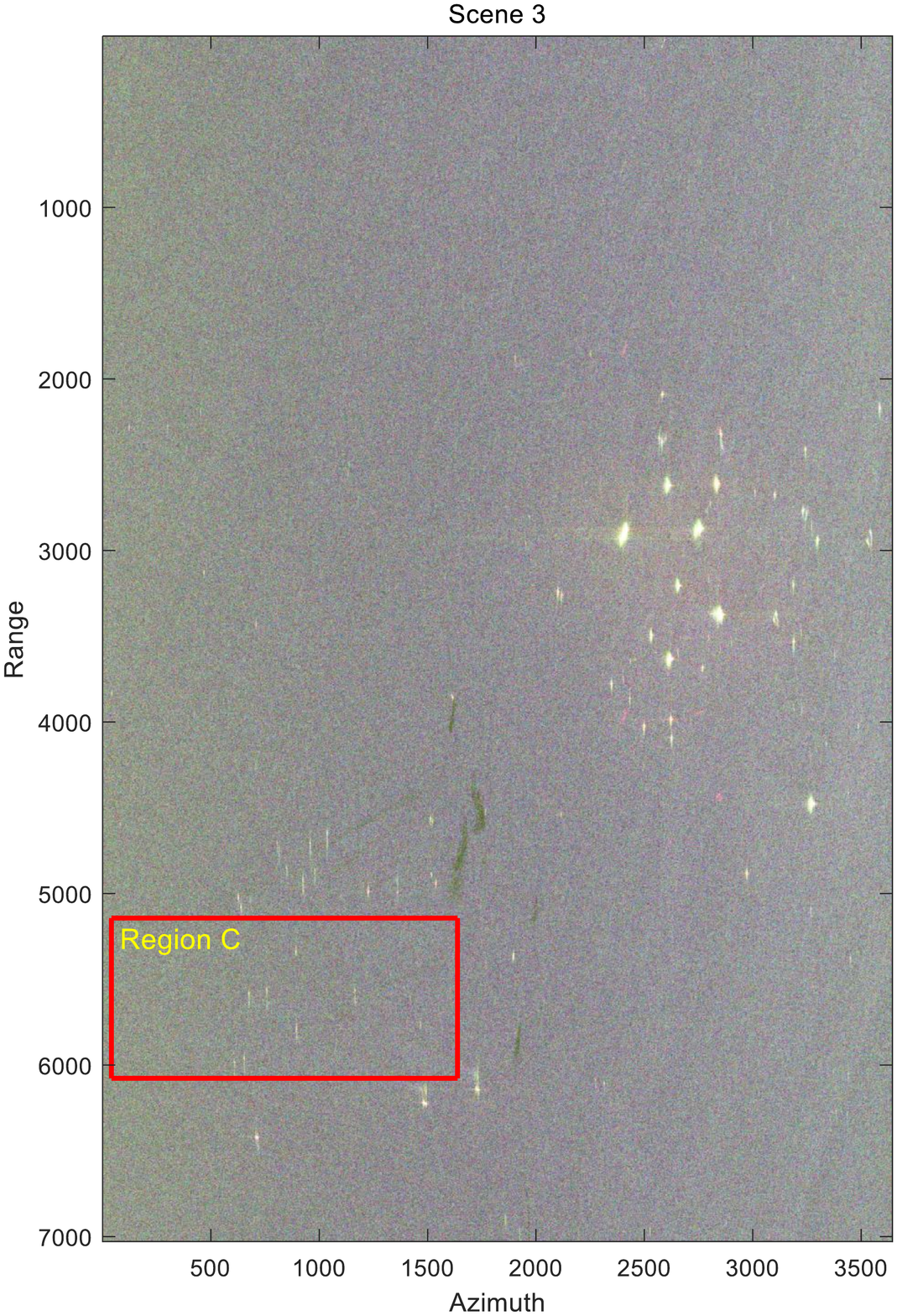}}
	\subcaptionbox{Scene 4}[0.32\linewidth]{\includegraphics[width=\linewidth]{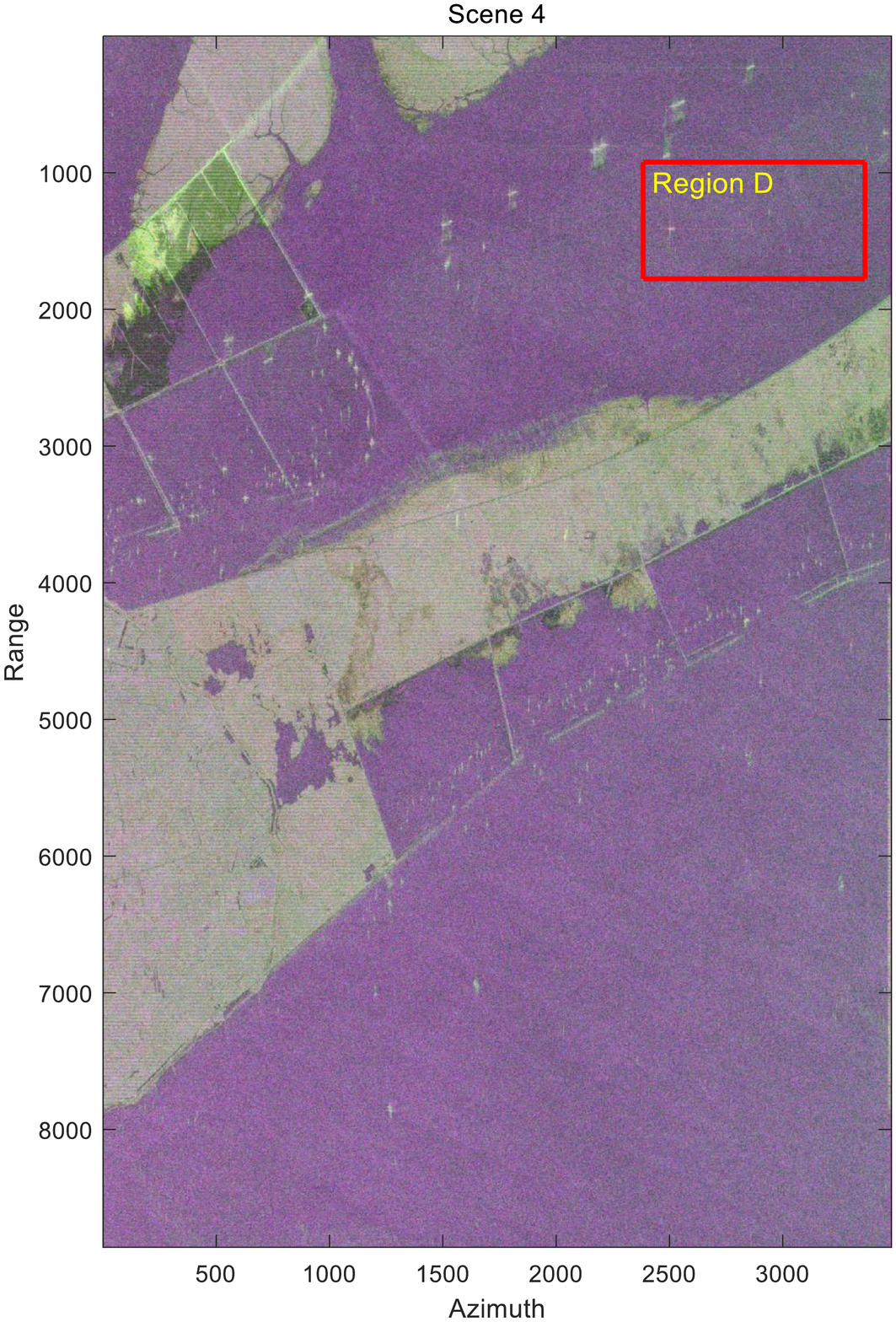}}
	\subcaptionbox{Scene 5}[0.32\linewidth]{\includegraphics[width=\linewidth]{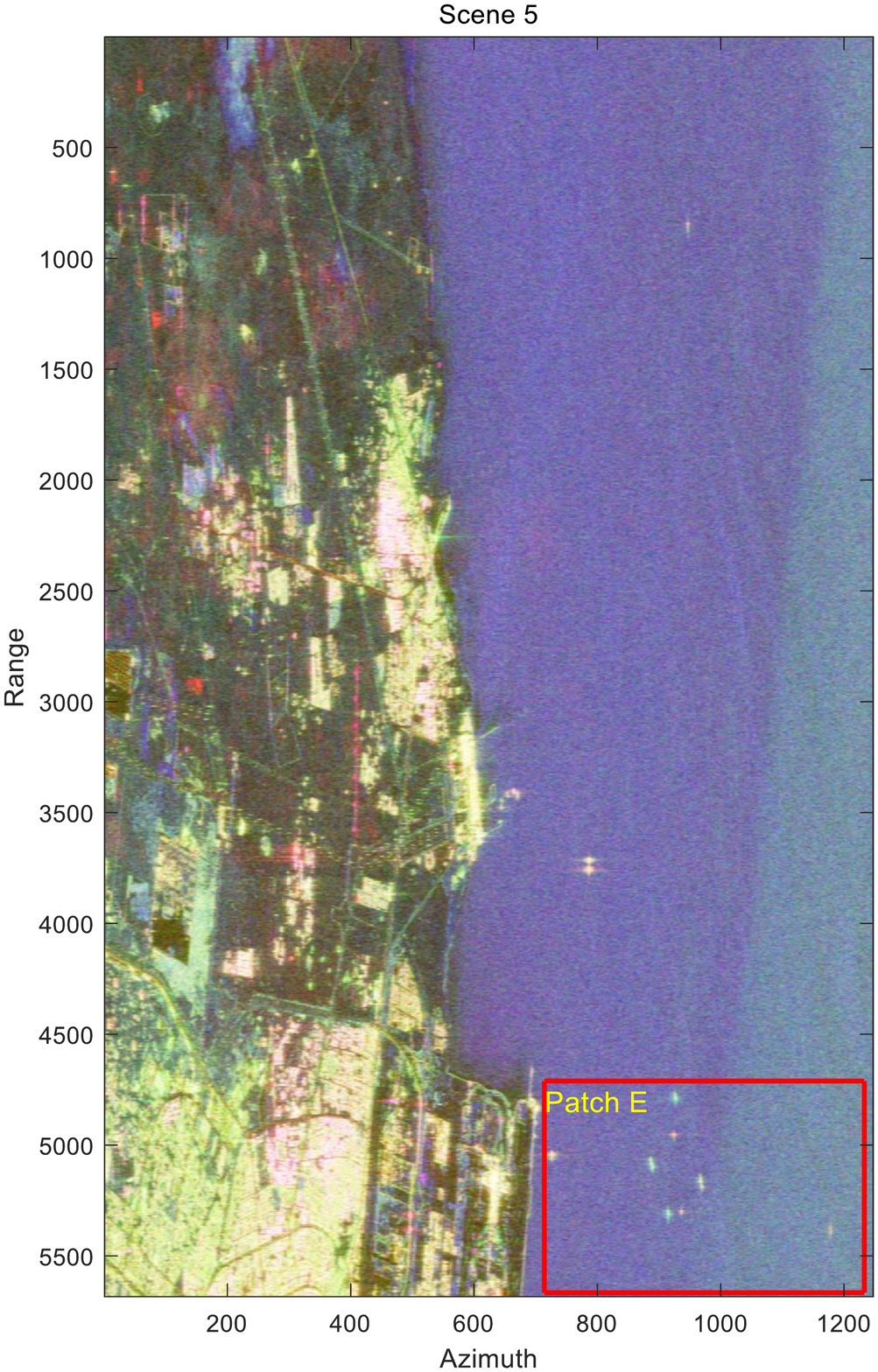}}
	\caption{Five experimental scenarios.}
	\label{fig:Scenes}
\end{figure}

The experiments are carried out on five SAR data sets, which were collected by using three types of space-borne platforms: the Canada C-band RADARSAT-2, the China C-band GaoFen-3, and the Japan L-band ALOS. Five typical scenes located in the South China Sea, Laizhou Bay, China, the Bohai Sea, China, the Yellow Sea, China, and the Persian Gulf, were selected for the experiments. All data products are fully polarimetric and in the single-look complex (SLC) format. The Pauli RGB images of scenes 1-5 are shown in Figure \ref{fig:Scenes} (a)-(e), where the color is coded by $|S_{HH} - S_{VV}|_{red}$, $|S_{HH} + S_{VV}|_{blue}$, $2|S_{HV}|_{green}$.\par

\begin{table}[h]
	\centering
	\caption{Detailed description of the datasets}
	\label{tab:dataInfo}
	\begin{tabular}{c|l|cccccc}
		\hline
		\multicolumn{2}{c|}{Scene ID}    & Product ID   &Sensor  &Date  &\tabincell{c}{Resolution \\R x A (m)}  & Incidence Angle($\textsuperscript{$\circ$}$) & $\beta$($\textsuperscript{$\circ$}$) \\ \hline
		\multicolumn{2}{c|}{1}           & RD2009000238-0  &RADARSAT-2  &2009/09/18 & 8.6 x 4.7 & 32.35\textsuperscript{$\circ$}-34.0\textsuperscript{$\circ$} &8.2\textsuperscript{$\circ$}\\ \hline
		\multicolumn{2}{c|}{2}           &PDS\_05350120 &RADARSAT-2   &2016/09/10  &6.8 x 5.1  &43.61\textsuperscript{$\circ$}-44.92\textsuperscript{$\circ$}   &10.2\textsuperscript{$\circ$}      \\ \hline
		\multicolumn{2}{c|}{3} &PDS\_03326690        & RADARSAT-2         &2011/09/08 &7.5 x 4.7 &42.77\textsuperscript{$\circ$}-44.11\textsuperscript{$\circ$} &10.7\textsuperscript{$\circ$}     \\ \hline
		\multicolumn{2}{c|}{4} &2075802        &GaoFen-3         &2016/11/20 &6.4 x 4.1 &25.27\textsuperscript{$\circ$}-27.80\textsuperscript{$\circ$} &2.1\textsuperscript{$\circ$}     \\ \hline
		\multicolumn{2}{c|}{5} &ALPSRP169460500        &ALOS         &2009/03/30 &3.5 x 7.7 &21.5\textsuperscript{$\circ$}-23.8\textsuperscript{$\circ$} &7.8\textsuperscript{$\circ$}     \\ \hline
	\end{tabular}
\end{table}
When evaluating the performance of different detectors, the ground-truth data is essential. However, for the five data sets in this paper, we don’t have ground-truth data, especially Automatic Identification System (AIS) data and meteorological data. Like the methods used in other papers \citep{zhang2018ship, zhang2020polsar, gao2018adaptive}, we determine the location of ships mainly through visual interpretation methods. This method is effective because the backscattered power from the ship is always stronger than the sea surface and ships appear as a bright spot on the image. In addition, the sea state corresponding to the test data (e.g., the roughness of sea surface) is also necessary for the evaluation of the performance of detectors. Therefore, we use the method in \citep{hajnsek2003inversion, zhang2019ship, yin2014extended} to assess the roughness of the sea surface of the test data, with the following equation
\begin{equation}
	sinc(4\beta_1) = \frac{T_{22}-T_{33}}{T_{22}+T_{33}}
\end{equation}
where $\beta_1 \in [0,90^{\circ}]$, the larger the $\beta_1$, the rougher the sea surface, and vice versa.
\subsection{Validation of G\textit{$\Gamma$}D with Different Detectors}
\label{s:val_GTD}
Statistical characterization of detectors is very important for the adaptive detection of detectors, such as the popular CFAR operation in the field of radar signal detection. In this paper, three known excellent detectors, DBSP, RsDVH, $\Delta S_n$ were chosen to show and verify the excepted effectiveness of the proposed detector joint-SA. The reason for the selection of DBSP, RsDVH and $\Delta S_n$ is based on the following facts. On the one hand, the three detectors are excellent for SAR image ship detection. Both  $\Delta S_n$, DBSP, and RsDVH have been proven to be effective detectors in characterizing target properties and ocean state and are superior to other detectors in improving the SCR, such as SPAN, PMS, and so on. On the other hand, detector DBSP and RsDVH have a good effect on the detection of relatively weakly scattering targets. Most importantly, it has been proven that the G$\Gamma$D  can effectively describe the statistical characterization of the two detectors. \par
\begin{figure}[H]
	
	\centering
	\includegraphics[width=\textwidth]{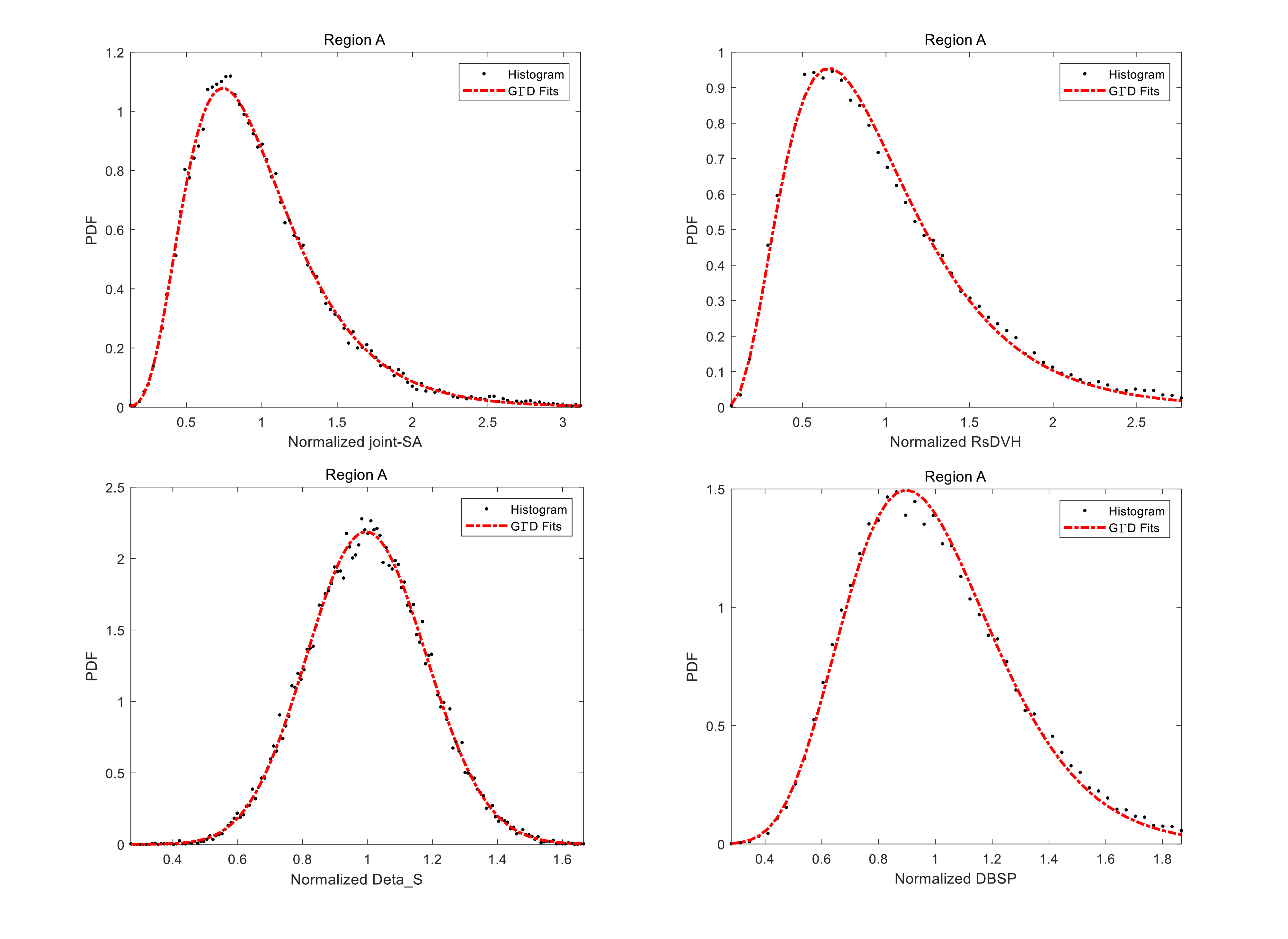}
	\caption{Histograms and the fit results of G$\Gamma$D for different detectors in Region A.}
	\label{fig:fit_Region_A}
\end{figure}

\begin{figure}[H]

	\centering
	\includegraphics[width=\textwidth]{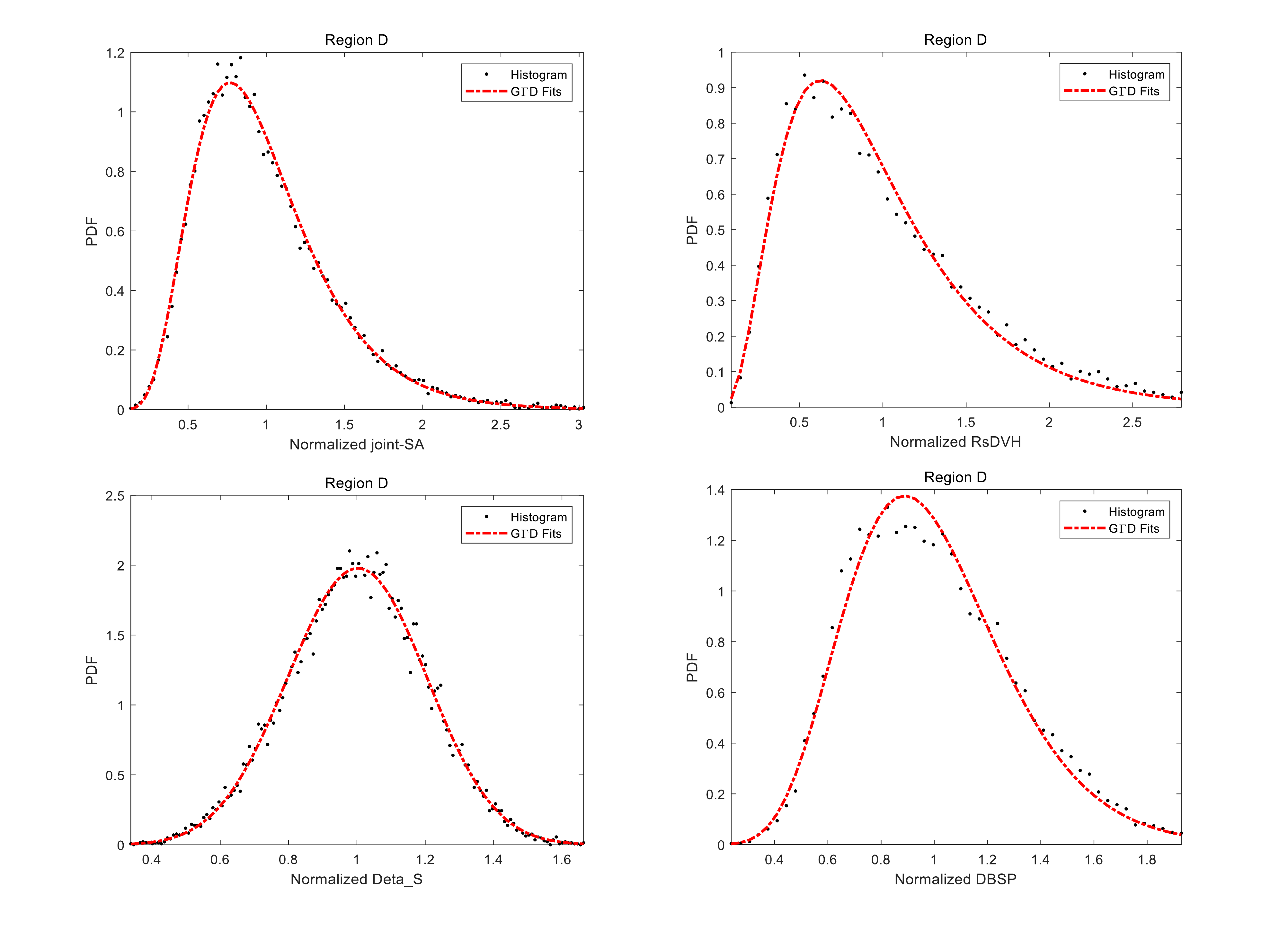}
	\caption{Histograms and the fit results of G$\Gamma$D for different detectors in Region D.}
	\label{fig:fit_Region_D}
\end{figure}

\begin{table}[h]
	
	\centering
	\caption{Quantitative evaluation of the fitting accuracy of five scenes using KL values for G$\Gamma$D}
	\label{tab:fit_accuracy}
	\begin{tabular}{c|l|ccccc}
		\hline
		\multicolumn{2}{c|}{Detectors}    & Patch A   &Patch B  &Patch C  &Patch D  & Patch E \\ \hline
		\multicolumn{2}{c|}{joint-SA}    & 0.0033   &0.0155  &0.0057  &0.0047  & 0.0064 \\ \hline
		\multicolumn{2}{c|}{RsDVH}    & 0.0041  &0.0072  &0.0064  &0.0086  & 0.0066 \\ \hline
		\multicolumn{2}{c|}{$\Delta$S}    & 0.0068  &0.0392  &0.0035  &0.0136  & 0.0050 \\ \hline
		\multicolumn{2}{c|}{DBSP}    & 0.0037  &0.0074  &0.0060  &0.0075  &0.0071 \\ \hline
	\end{tabular}
\end{table}
To be fair, in this paper we also choose the G$\Gamma$D to describe the statistical characterization of the proposed detector. The initial intention of doing so is based on the following two considerations: First, compared to other classic statistical models, such as Gaussian, Rayleigh, Gamma, log-normal, Weibull, and K, many results were shown that G$\Gamma$D can obtain higher fitting accuracy for single-channel sea clutter data. Second, G$\Gamma$D is a more advanced and general statistical model, because many statistical models in the field of SAR images are a special case of G$\Gamma$D in theory.\par

\begin{figure}[H]
	\centering
	\includegraphics[width=\textwidth]{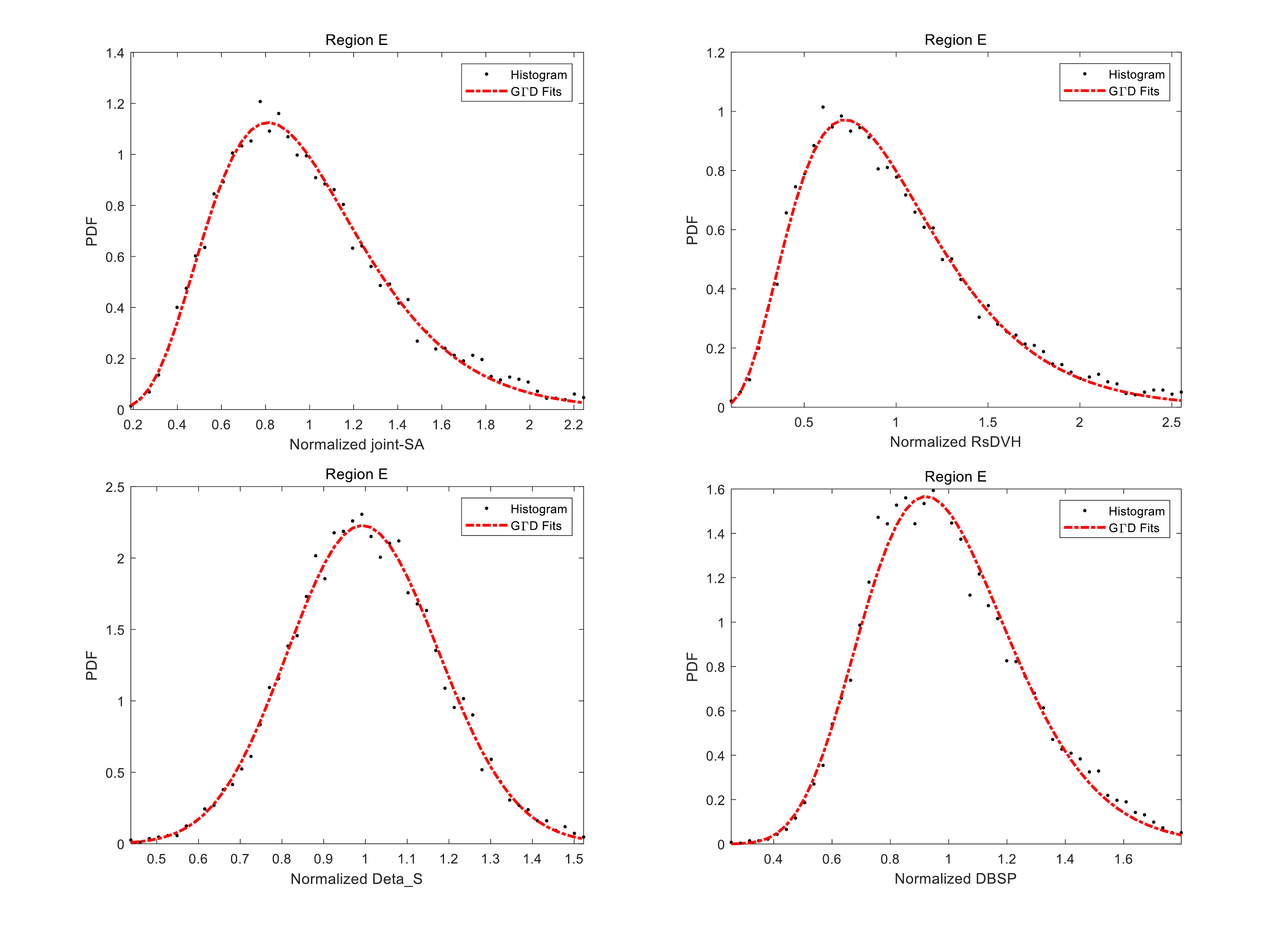}
	\caption{Histograms and the fit results of G$\Gamma$D for different detectors in Region E.}
	\label{fig:fit_Region_E}
\end{figure}
For each region in the scene, we used G$\Gamma$D  to fit each of the $\Delta S_n$, DBSP, RsDVH, joint-SA. Meanwhile, to make the fitting results clearer and more intuitive, we normalized all the data before fitting them by dividing them by the mean of the corresponding data. In addition, to better illustrate the fitting accuracy of G$\Gamma$D, the commonly symmetric Kullback-Leibler (KL)\citep{kullback1951information} distance was introduced, being used as the measure of the difference in similarity between the estimated and theoretical pdfs.\par

As shown in figures \ref{fig:fit_Region_A}, \ref{fig:fit_Region_D}, \ref{fig:fit_Region_E} and table \ref{tab:fit_accuracy}, the G$\Gamma$D show a good-of-fit for different detectors, and the fitting result could not be affected by external factors, such as radar sensor type, sea surface roughness, etc. Therefore, the results of fitting experiments of the G$\Gamma$D  are acceptable, and the G$\Gamma$D  can be used as a statistical model for adaptive threshold estimation. Because of the same results and to save space, we only show the fitting results for different satellite platforms and different incidence angles.

\subsection{Effectiveness validation of the joint-SA detector}
\label{s:eff_val_detector}
After completing the validation of the G$\Gamma$D for different detectors, we will focus on the effectiveness of the proposed detector for relatively weakly scattering targets. In the next pages, we will compare the ability of different detectors to highlight relatively weakly scattering targets. In the field of SAR image ship detection, the SCR has been regarded as a core physical quantity to evaluate the goodness of a detector. Therefore, we also used the SCR to evaluate detectors mentioned in this paper. \par
The first thing we must do is select a region that contains relatively weakly scattering targets. A flexible way to process images is to select a chip from scenes that contain only one relatively weakly scattering target and its local background. The initial intention of doing this is based on two considerations. On the one hand,  from the perspective of target highlighting, the  region containing only one target can show the state of the target and the background more intuitively and ensure that the target is not affected by other factors. On the other hand, from the visual point of view, it is possible to explore and reveal the details of different detectors more clearly in a limited space. It should be noted that due to the lack of AIS data, all ships in the paper were visually gained from images by trained image interpreters.\par
\begin{figure}[H]
	\centering
	\includegraphics[width=\textwidth]{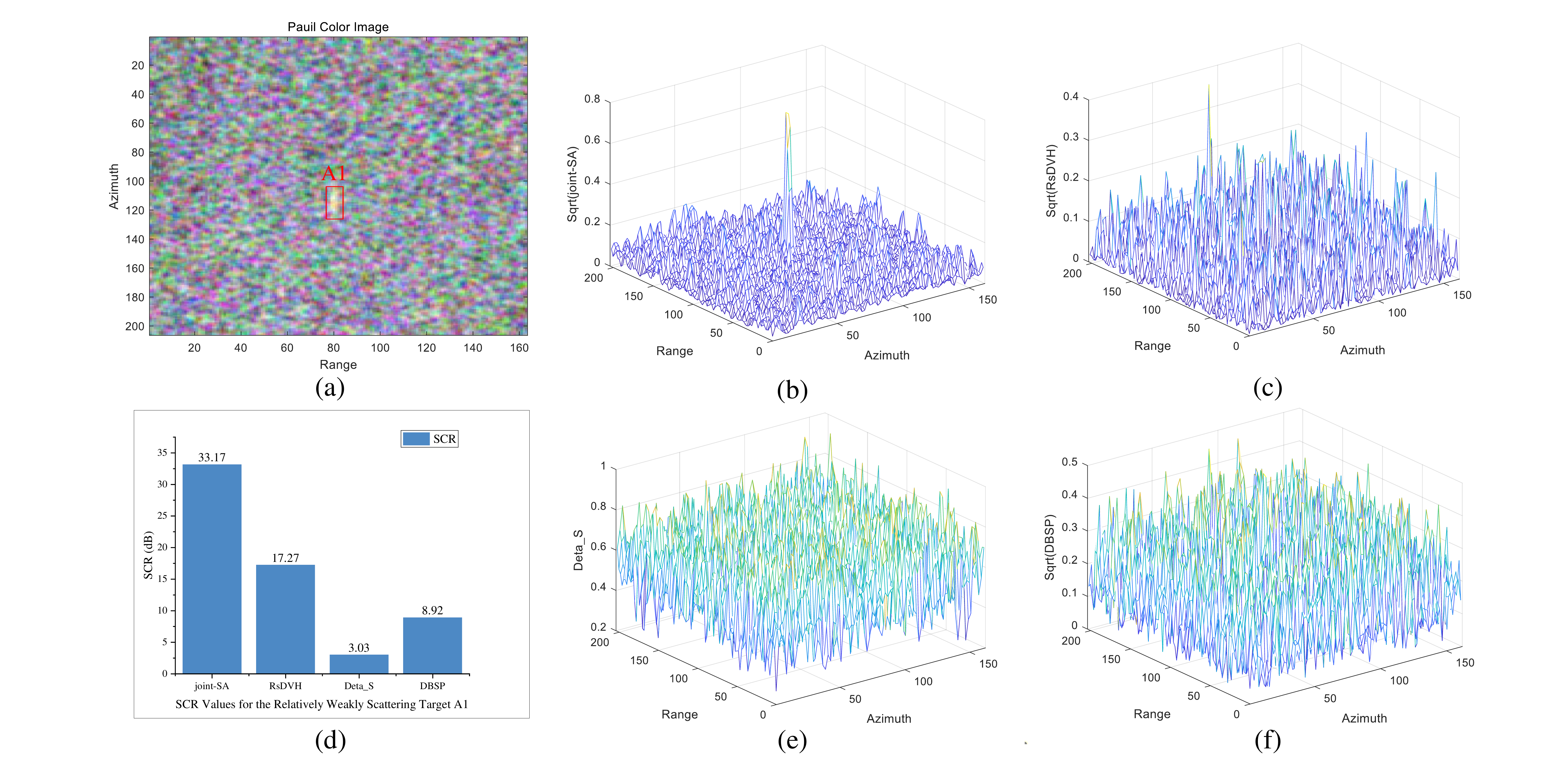}
	\caption{Comparison of different detectors of a relatively weakly scattering ship on region A in scene 1. (a) Pauli RGB image. (b) 3-D display of the joint-SA for the interval [0,1]. (c) 3-D display of the RsDVH for the interval [0,1]. (d) SCR value for the target A1. (e) 3-D display of the $\Delta$S for the interval [0,1]. (f) 3-D display of the DBSP for the interval [0,1].}
	\label{mesh:Region_A}
\end{figure}

First, we performed qualitative analysis and quantitative evaluation on the relatively weakly scattering target from region A in scene 1, which is related to C-band RADARSAT-2 and the incidence angle of scene 1 is [32.35°, 34.01°]. As figure \ref{mesh:Region_A} showed, (a) is the Pauli image with a relatively weakly scattering target, and (b), (c), (e), (f) is the 3-D image with different detectors. As can we see, joint-SA performs best in target highlighting and clutter suppression, while RsDVH, $\Delta$S, and DBSP are unable to distinguish the target. To further illustrate the effectiveness of joint-SA, we compared the SCR values of the four detectors, as shown in figure \ref{mesh:Region_A} (d). The detector joint-SA has the highest SCR value of 33.17 (dB), which is twice as high as the detector RsDVH, 10 times as high as  $\Delta$S, and 4 times as high as DBSP. Therefore, from both qualitative analysis and quantitative evaluation, we can conclude that joint-SA is better than the other detectors in this group of experiments.\par

\begin{figure}[H]
	\centering
	\includegraphics[width=\textwidth]{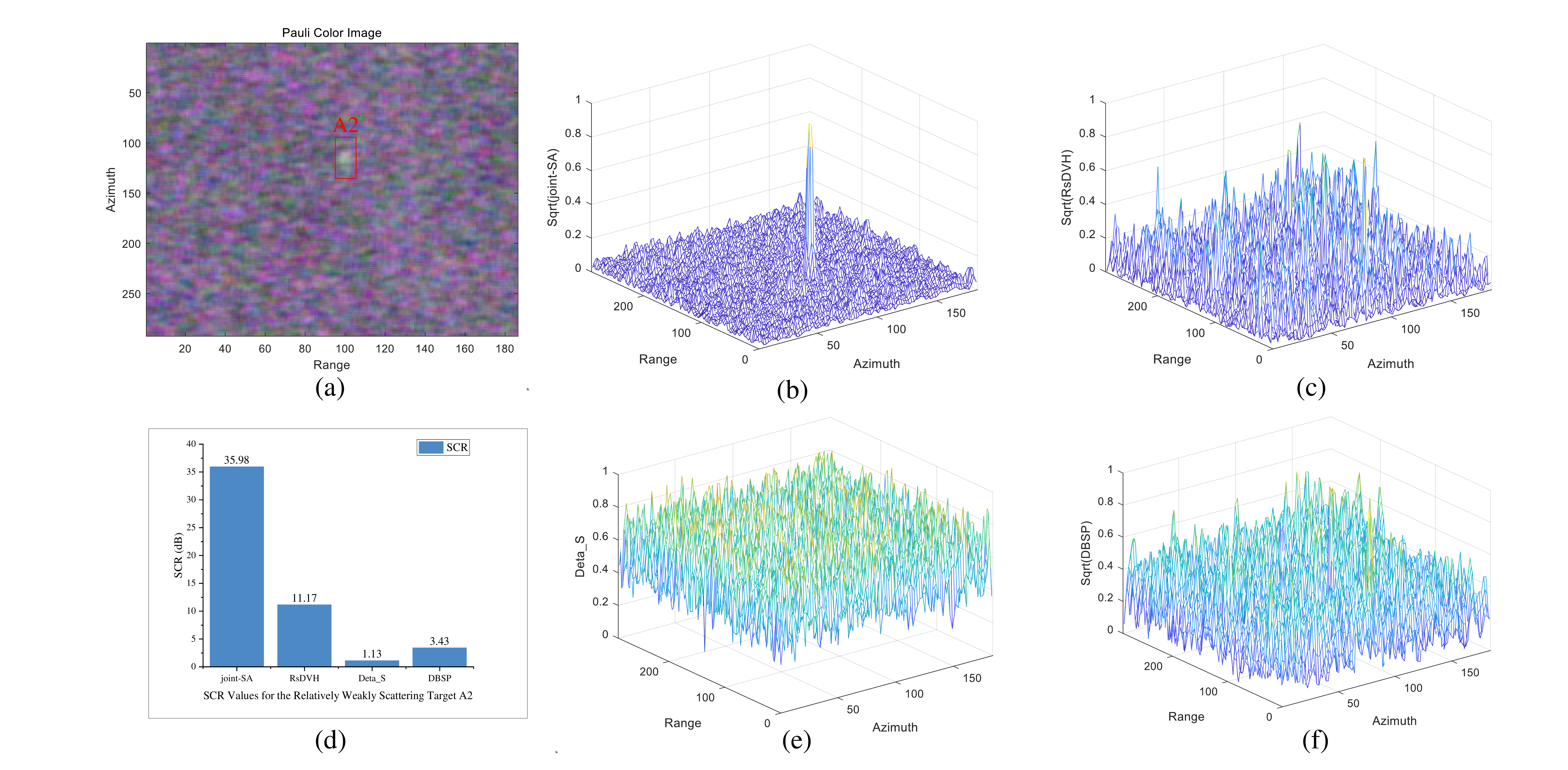}
	\caption{Comparison of different detectors of a relatively weakly scattering ship on region D in scene 4. (a) Pauli RGB image. (b) 3-D display of the joint-SA for the interval [0,1]. (c) 3-D display of the RsDVH for the interval [0,1]. (d) SCR value for the target A2. (e) 3-D display of the $\Delta$S for the interval [0,1]. (f) 3-D display of the DBSP for the interval [0,1].}
	\label{mesh:Region_D}
\end{figure}

Second, we performed qualitative analysis and quantitative evaluation on the target from region D in scene 4, which is related to C-band GF-3, and the incidence angle of scene 4 is [25.27°, 27.80°]. As shown in figure \ref{mesh:Region_D}, like the first set of experiments, the detector joint-SA outperforms the other three detectors. As can be seen in figure \ref{mesh:Region_D} (b), joint-SA has a very good ability to suppress the clutter and highlight the target. In addition, the SCR value of the detector joint-SA is higher than this of the other three detectors, as can be seen in figure \ref{mesh:Region_D} (d). This is consistent with the 3-D visualization results plot. Hence, the detector joint-SA is also better than the other detectors in this set of experiments.\par

\begin{figure}[H]
	\centering
	\includegraphics[width=\textwidth]{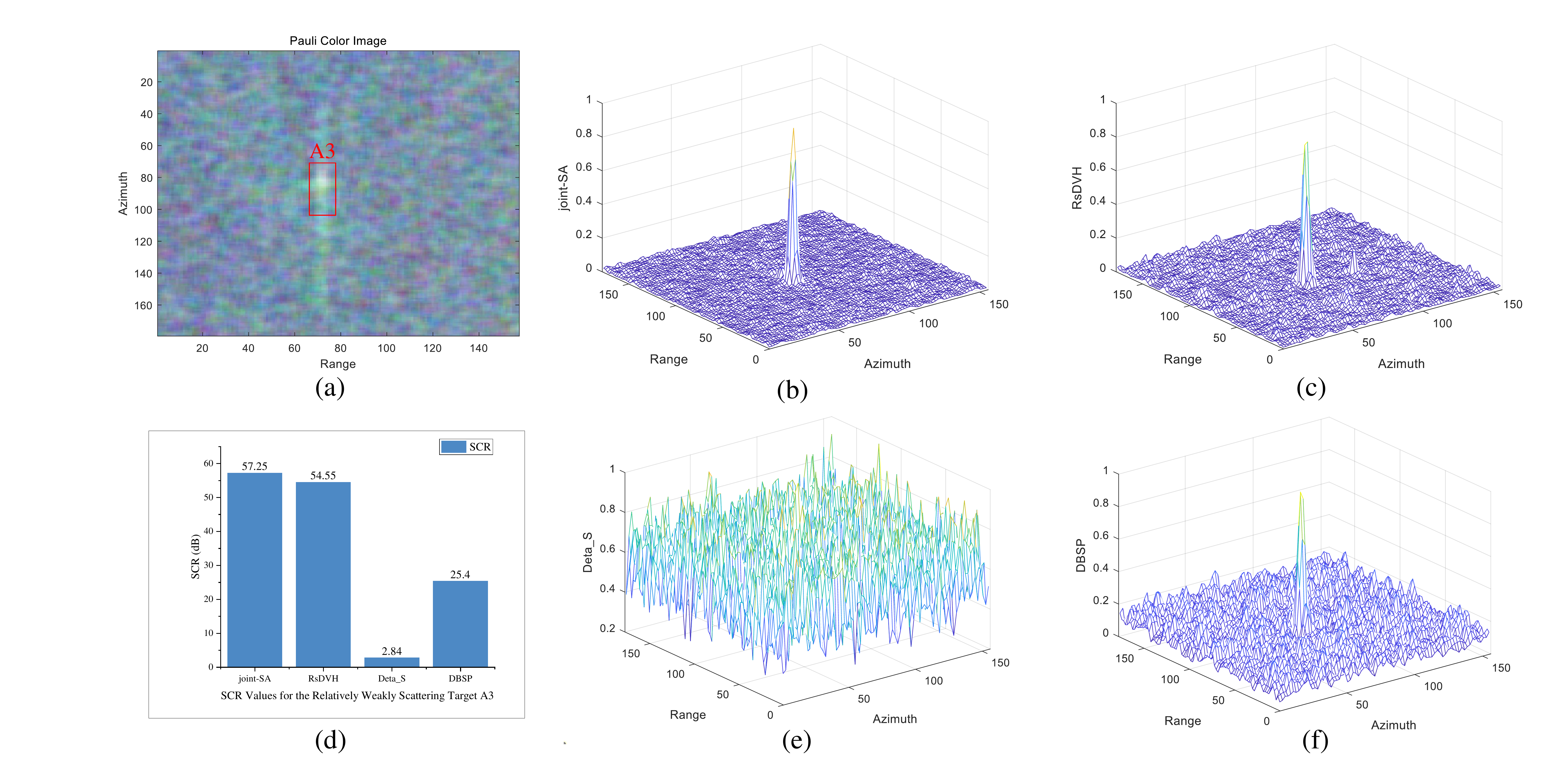}
	\caption{Comparison of different detectors of a relatively weakly scattering ship on region E in scene 5. (a) Pauli RGB image. (b) 3-D display of the joint-SA for the interval [0,1]. (c) 3-D display of the RsDVH for the interval [0,1]. (d) SCR value for the target A3. (e) 3-D display of the $\Delta$S for the interval [0,1]. (f) 3-D display of the DBSP for the interval [0,1].}
	\label{mesh:Region_E}
\end{figure}
Third, we performed qualitative analysis and quantitative evaluation on the target from region E in scene 5, which is related to L-band ALOS. The incidence angle of scene 5 is [21.5°, 23.8°]. As shown in figure \ref{mesh:Region_E}, except for the detector $\Delta$S, all three detectors can easily distinguish the targets, and the peaks of the three detectors are similar. But for the DBSP, the clutter rejection is worse than that of joint-SA and RsDVH. To compare the ability of these three detectors to highlight the target, we also compare their SCR values, as shown in figure \ref{mesh:Region_E} (d), the SCR value of the three detectors is 57.25 dB, 54.55 dB, and 25.40 dB, which are consistent with the performance in the 3-D images. Therefore, the detector joint-SA is also better than the other detectors in this set of experiments.\par

Because of the same results, region B in scene 2 and region C in scene 3 which is related to the RADARSAT-2 in C-band type, we do not analyze in detail. Through the above three sets of experiments, we can conclude that the four detectors have different sensitivities to clutter pixels and target pixels in different regions, and the proposed ship detector joint-SA is better than any of the detectors compared in this paper, both in terms of target prominence, clutter rejection, and SCR values. In addition, we can also see that in the scenes with low incidence angles, detectors joint-SA, RsDVH and DBSP have good performance results, but with the increasing incidence angles, the effectiveness of detectors RsDVH and DBSP is decreasing, while for the joint-SA, the effectiveness does not change with the incidence angle when the incidence angle is less than 45°.

\subsection{Comparison of CFAR results with different detectors}
In section \ref{s:val_GTD}, we have verified the G$\Gamma$D  can be used to describe the statistics of the four detectors, which provides the possibility to subsequently compare the CFAR detection outperformance of the four detectors. In section \ref{s:eff_val_detector}, we also demonstrated that the detector joint-SA outperforms $\Delta$S, DBSP, and RsDVH in SCR. Therefore, we can conduce that the the proposed detector in this paper still outperforms the other three detectors in CFAR adaptive detection. Next, we will prove our point with experiments. It should be noted that the $p_{fa}$  is$ 10^{-5}$ in this paper.\par

\begin{figure}[H]
	\centering
	\includegraphics[width=\textwidth]{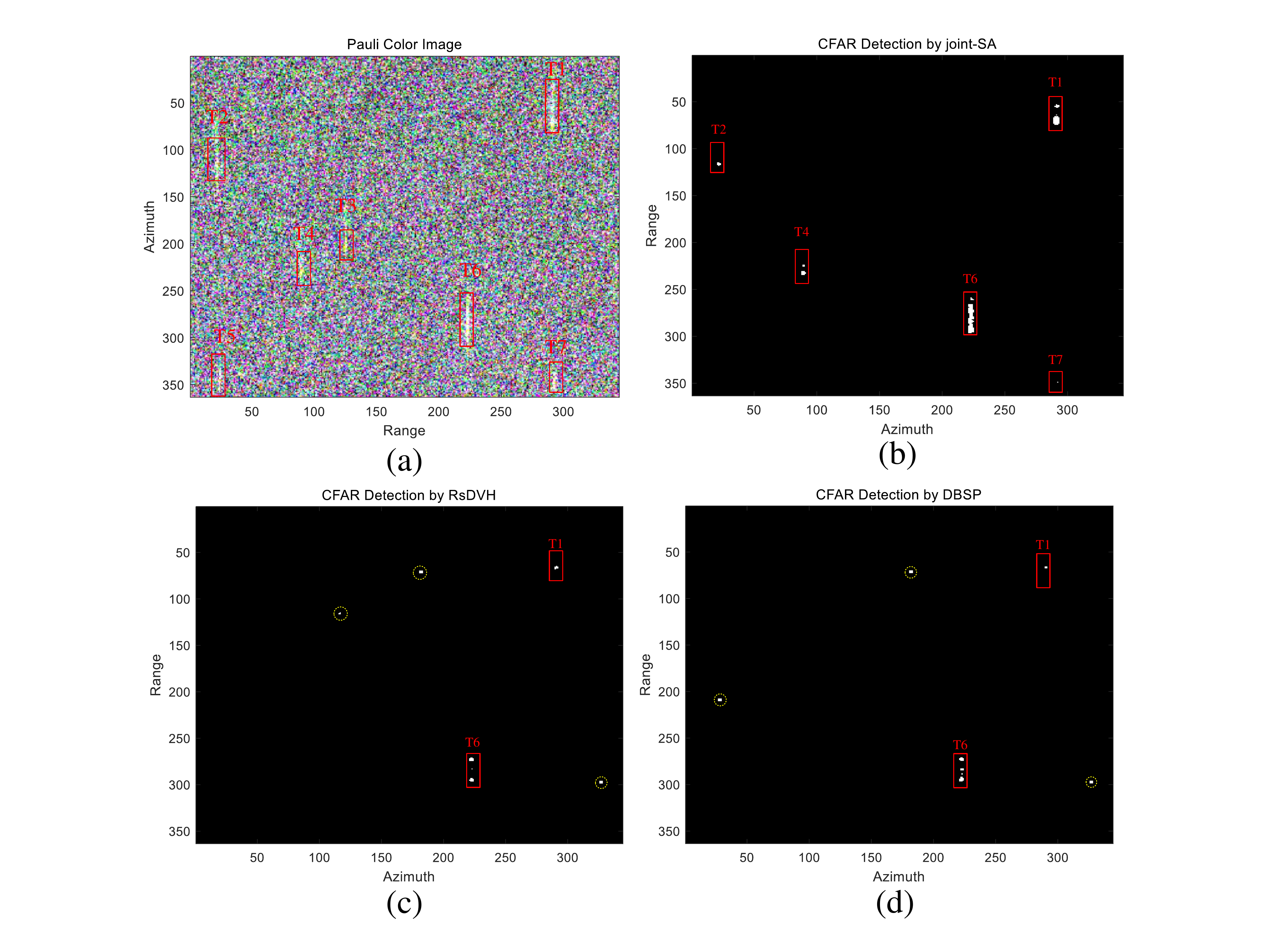}
	\caption{Detection results of region A in scene 1. (a) Pauli RGB image. (b) CFAR detection result by joint-SA. (c) CFAR detection 
	result by RsDVH. (d) CFAR detection result by DBSP. Yellow dashed circles: false-alarm targets, and the red boxes: targets}
    \label{Fig:det_Region_A}
\end{figure}

\begin{figure}[H]

	\centering
	\includegraphics[width=\textwidth]{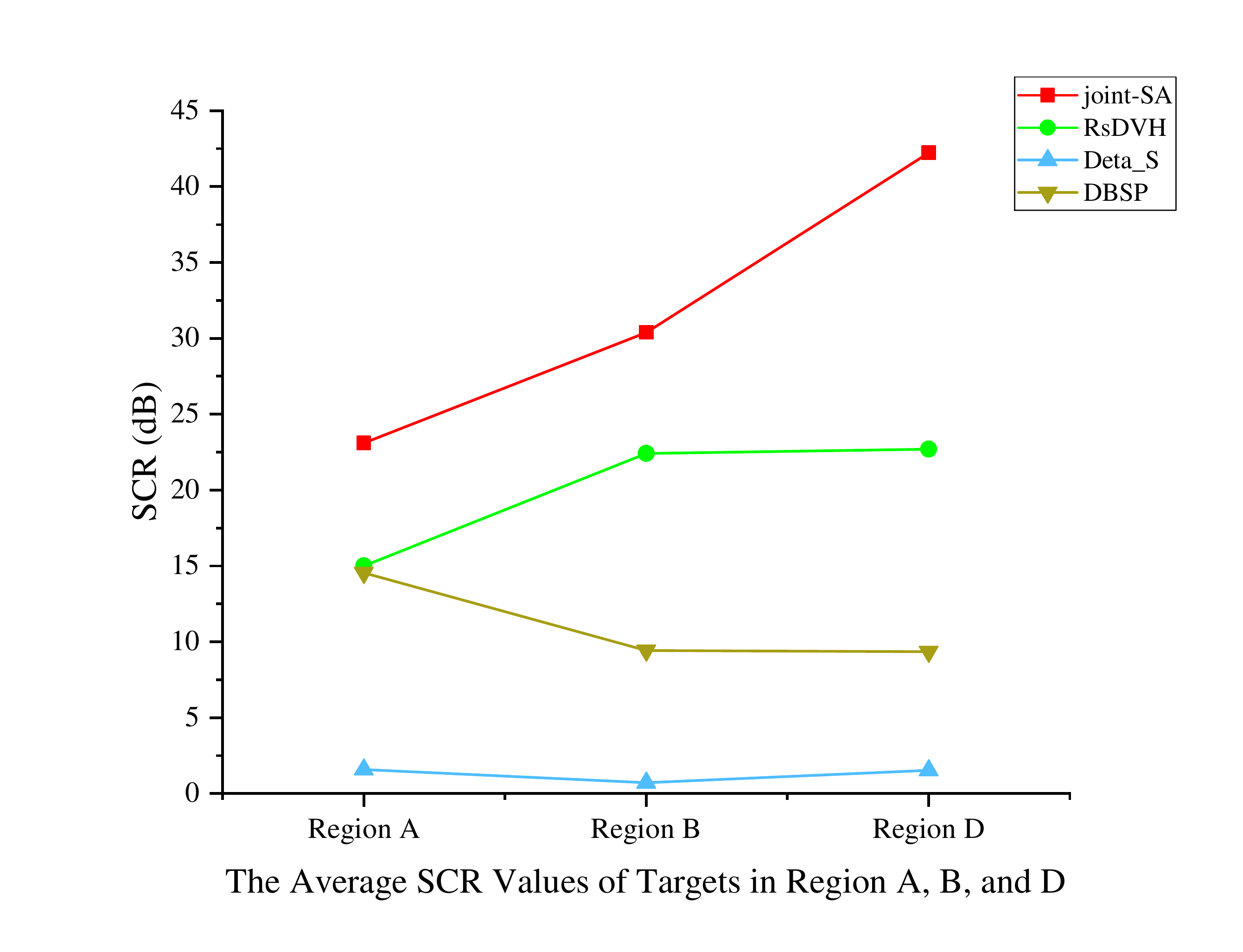}
	\caption{The average SCR values of targets in regions A, B, and D.}
	\label{fig:det_avg_SCR}
\end{figure}

The first experiment will use region A in scene 1. The incidence angle of scene 1 is [32.35°, 34.01°]. As can be seen in figure \ref{Fig:det_Region_A} (a), there are seven weakly scattering ships. The detector joint-SA detected five targets without false alarms. While for the RsDVH and DBSP, they detected only two targets, but with three false alarms. Therefore, from the perspective of detection accuracy, joint-SA is superior to the other three detectors. As shown in figure \ref{fig:det_avg_SCR}, the average SCR value of joint-SA is also higher than the other three detectors. To further analyze the detection performance of the four detectors, we counted the SCR values of them. As can be seen in table \ref{tab:A_SCR}, the SCR values of joint-SA of T3 and T5 are 12.92 dB and 13.03 dB, respectively, which are the lowest among all targets. For both RsDVH and DBSP, the undetected targets are those with low SCR values, and surprisingly, the SCR value of target T2 appears negative in both detectors. Therefore, joint-SA is superior to the other four detectors in terms of detection accuracy and SCR.\par
\begin{table}[h]
	
	\centering
	\caption{The SCR values for relatively weakly scattering targets in region A}
	\label{tab:A_SCR}
	
	\begin{tabular}{c|l|ccccccc}
		\hline
		\multicolumn{2}{c|}{Detectors}    & T1   &T2  &T3  &T4  & T5 &T6 &T7\\ \hline
		\multicolumn{2}{c|}{RsDVH}    & 25.95   &-0.67  &4.75  &22.59  &10.01 &26.28 &21.08 \\ \hline
		\multicolumn{2}{c|}{$\Delta$S}    &0.69  &-0.35  &2.53  &2.46  &1.80 &1.70 &2.32 \\ \hline
		\multicolumn{2}{c|}{DBSP}    & 20.97  &-0.47  &2.72  &20.12  & 9.99 &26.71 &21.64 \\ \hline
		\multicolumn{2}{c|}{\textbf{joint-SA}}    &\textbf{38.65}  &\textbf{13.88}  &\textbf{12.92}  &\textbf{24.41}  &\textbf{13.03} &\textbf{42.15} &\textbf{46.66} \\ \hline
	\end{tabular}
\end{table}

\begin{figure}[H]
	
	\centering
	\includegraphics[width=\textwidth]{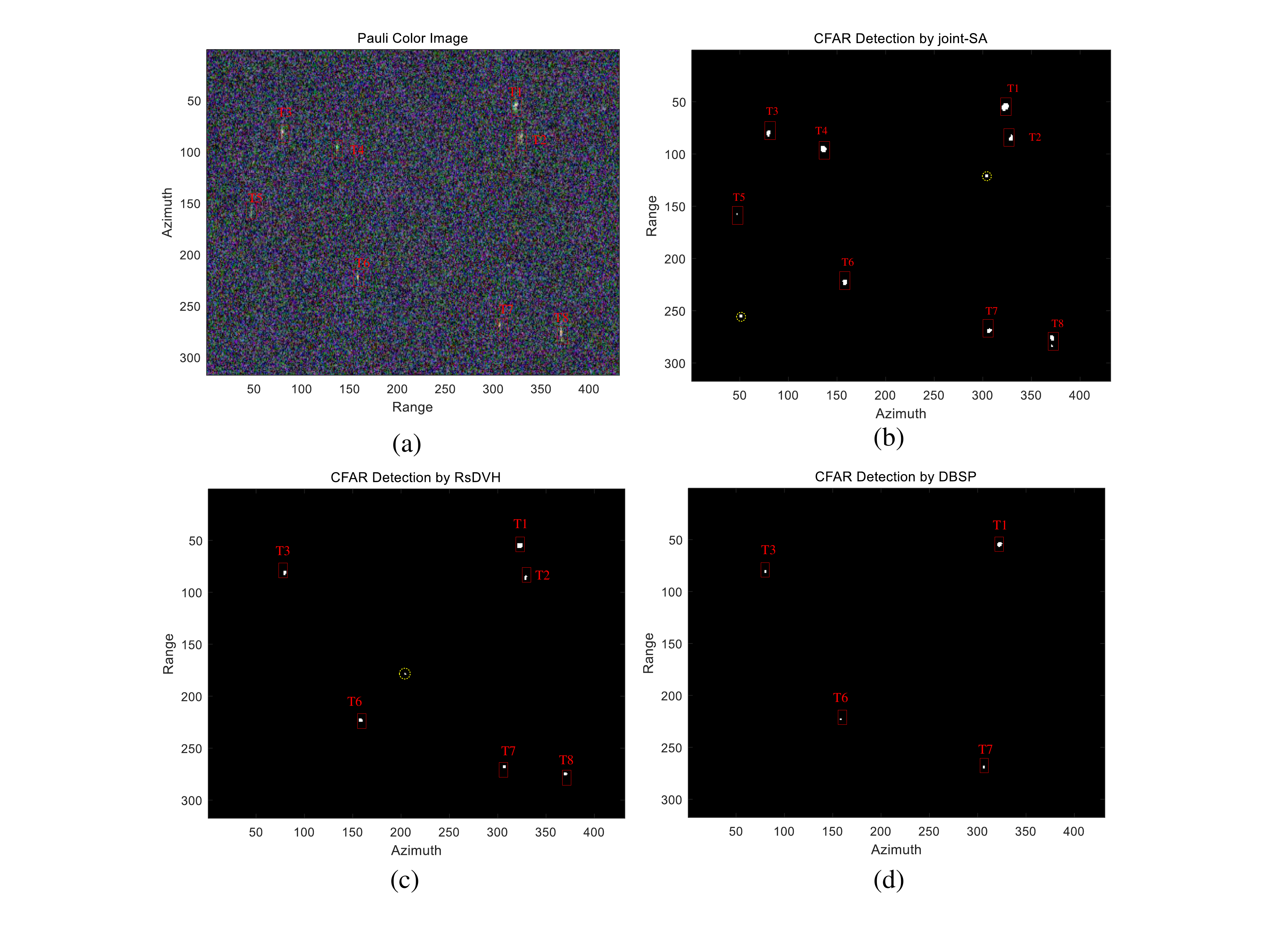}
	\caption{Detection results of region B in scene 2. (a) Pauli RGB image. (b) CFAR detection result by joint-SA. (c) CFAR detection result by RsDVH. (d) CFAR detection result by DBSP. Yellow dashed circles: false-alarm targets, and the red boxes: targets.}
	\label{Fig:det_Region_B}
\end{figure}

The second set of experiments will be conducted on region B scene 2 which is related to the C-band RADARSAT-2. The incidence angle of scene 2 is [43.61°, 44.92°]. As can be seen from figure \ref{Fig:det_Region_B}, there are 8 relatively weakly scattering targets in this scene. In terms of the detection accuracy, joint-SA successfully detected all targets, while RsDVH and DBSP detected 6, and 4 targets, respectively. To further evaluate the performance of the four detectors, we also counted the average SCR values of the eight targets. It can also be seen from figure \ref{fig:det_avg_SCR} that the average SCR value for the 8 targets, joint-SA is the highest among all detectors. In table \ref{tab:B_SCR}, targets T4 and T5 were missed with SCR values of 9.21 dB and 8.97 dB in RsDVH, respectively, with the lowest SCR values among RsDVH, and the difference in SCR value between T5 (8.97 dB) and T1 (40.44 dB) is about 5 times. A similar situation occurred in the DBSP, where targets with lower SCR values were not detected, and the difference between T4 (3.37 dB) and T1 (17.01 dB) is 5 times. It is the large difference in SCR values that leads to incorrect estimation in CFAR adaptive threshold estimation and thus targets miss.\par
\begin{figure}[H]

\centering
\includegraphics[width=\textwidth]{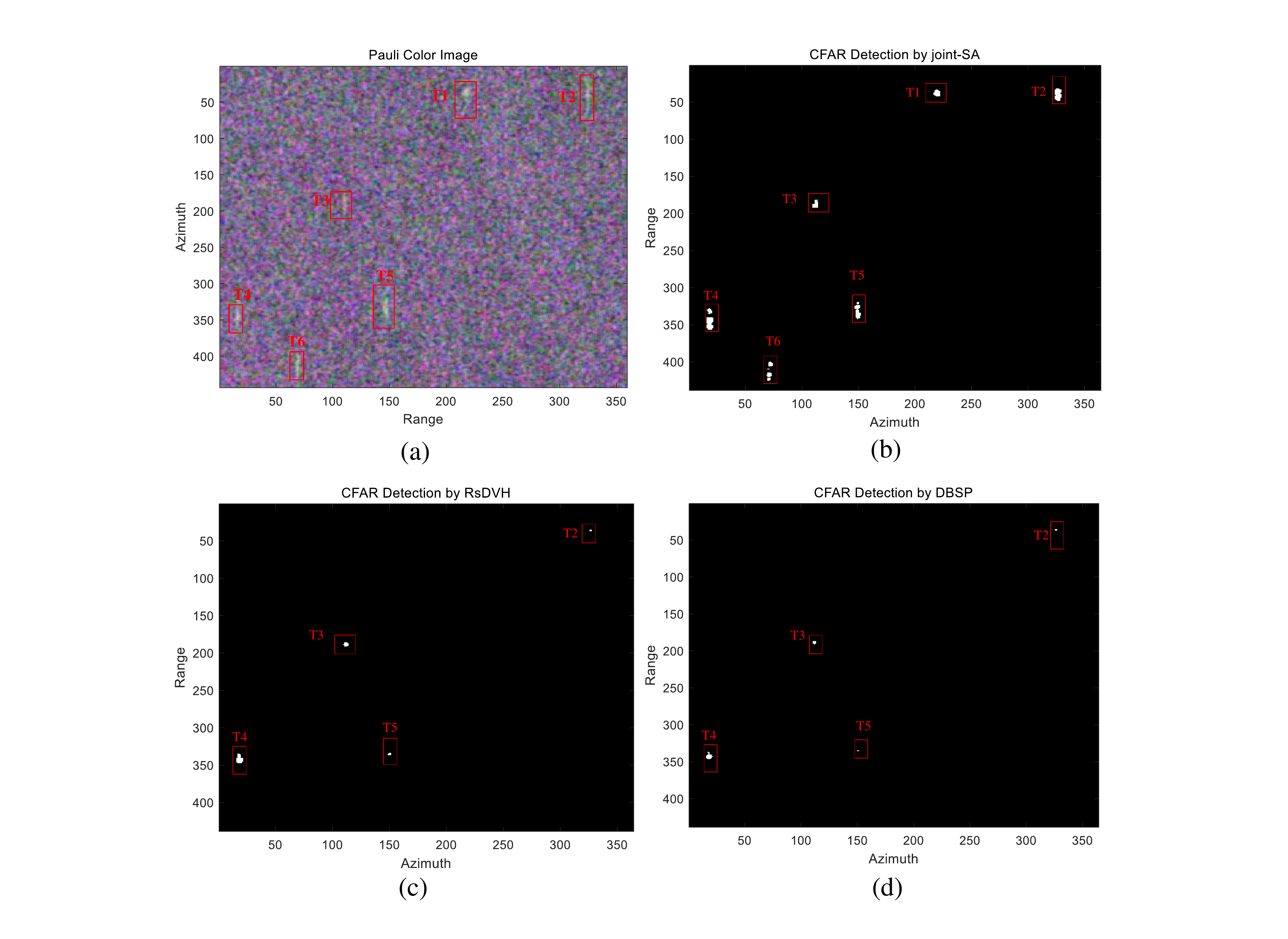}
\caption{Detection results of region D in scene 4. (a) Pauli RGB image. (b) CFAR detection result by joint-SA. (c) CFAR detection result by RsDVH. (d) CFAR detection result by DBSP. Yellow dashed circles: false-alarm targets, and the red boxes: targets.}
\label{Fig:det_Region_D}
\end{figure}

\begin{table}[h]
	
	\centering
	\caption{The SCR values for relatively weakly scattering targets in region B}
	\label{tab:B_SCR}
	
	\begin{tabular}{c|l|cccccccc}
		\hline
		\multicolumn{2}{c|}{Detectors}    & T1   &T2  &T3  &T4  & T5 &T6 &T7 &T8\\ \hline
		\multicolumn{2}{c|}{RsDVH}    & 40.44   &25.42  &22.43  &9.21  &8.97 &24.02 &23.94&24.78 \\ \hline
		\multicolumn{2}{c|}{$\Delta$S}&0.76  &1.47  &1.82  &0.41  &0.44 &-0.04 &1.32& -0.48\\ \hline
		\multicolumn{2}{c|}{DBSP}    & 17.01  &11.1  &8.01  &3.37  & 4.05 &9.49 &12.79& 9.63\\ \hline
		\multicolumn{2}{c|}{\textbf{joint-SA}}    &\textbf{48.77}  &\textbf{27.95}  &\textbf{28.22}  &\textbf{40.68}  &\textbf{17.84} &\textbf{25.56} &\textbf{26.03}&\textbf{27.11} \\ \hline
	\end{tabular}
\end{table}
The third experiment will use region D in scene 4. The incidence angle is [25.27°, 27.80°]. As can be seen in figure\ref{Fig:det_Region_D} (a), there are six relatively weakly scattering targets in this region. From the (b), (c), (d), we can see that the detector joint-SA successfully detected 6 targets, while detectors RsDVH and DBSP only detected 4 targets. In terms of detection accuracy, joint-SA is better than RsDVH and DBSP. It can also be seen from figure \ref{fig:det_avg_SCR} that the average SCR value of joint-SA is much larger than those of RsDVH and DBSP. Therefore, both in terms of detection accuracy and SCR values, joint-SA is better than the other detectors.\par

\begin{table}[h]
	
	\centering
	\caption{The SCR values for relatively weakly scattering targets in region D}
	\label{tab:D_SCR}
	
	\begin{tabular}{c|l|ccccccc}
		\hline
		\multicolumn{2}{c|}{Detectors}    & T1   &T2  &T3  &T4  & T5 &T6 \\ \hline
		\multicolumn{2}{c|}{RsDVH}    & 14.09   &26.67 &28.97  &42.46  &18.99 &4.91  \\ \hline
		\multicolumn{2}{c|}{$\Delta$S}    &2.48  &1.43  &1.86  &1.35  &1.15 &0.86 \\ \hline
		\multicolumn{2}{c|}{DBSP}    & 4.74  &14.18  &9.93  &17.79  & 6.53 &2.94  \\ \hline
		\multicolumn{2}{c|}{\textbf{joint-SA}}    &\textbf{37.67}  &\textbf{55.73}  &\textbf{33.38}  &\textbf{54.87}  &\textbf{37.81} &\textbf{33.92} \\ \hline
	\end{tabular}
\end{table}

Similarly, we counted the SCR of six relatively weakly scattering targets in different detectors. As can be seen in table \ref{tab:D_SCR}, the joint-SA has higher SCR values than the other four detectors. For the detectors joint-SA, the difference in SCR values for individual targets is small, while the difference in SCR values for individual targets in RsDVH and DBSP is large. We can see that the SCR values of T1 and T6 are 14.09 dB and 4.91 dB, respectively, which are the smallest SCR values among all targets, and the largest SCR value T4 is 42.46, which is 9 times different from the smallest T6. It is the huge difference in SCR values between different targets that leads to the inability to estimate the threshold correctly in the CFAR adaptive detection stage, thus making T1 and T6 missed detections, and the same reason for the detector DBSP. \par
In our experiments, we found that some targets with low average SCR values were detected, while some targets with high average SCR values were not detected. This phenomenon is reasonable for the following reasons: on the one hand, it should be noted that since we can’t know all pixels of targets accurately, we calculate the SCR value by boxing the target and the clutter region, average the values in the region, and then calculating the SCR. Therefore, a high average SCR does not mean that every target pixel has a high SCR value, but vice versa. On the other hand, when a target with a low average SCR value has a pixel with a high SCR value, the target will also be detected, and conversely, when the target has high average SCR values, but every single pixel is slightly lower, the target will not be detected.\par
\begin{figure}[H]
	\centering
	\includegraphics[width=\textwidth]{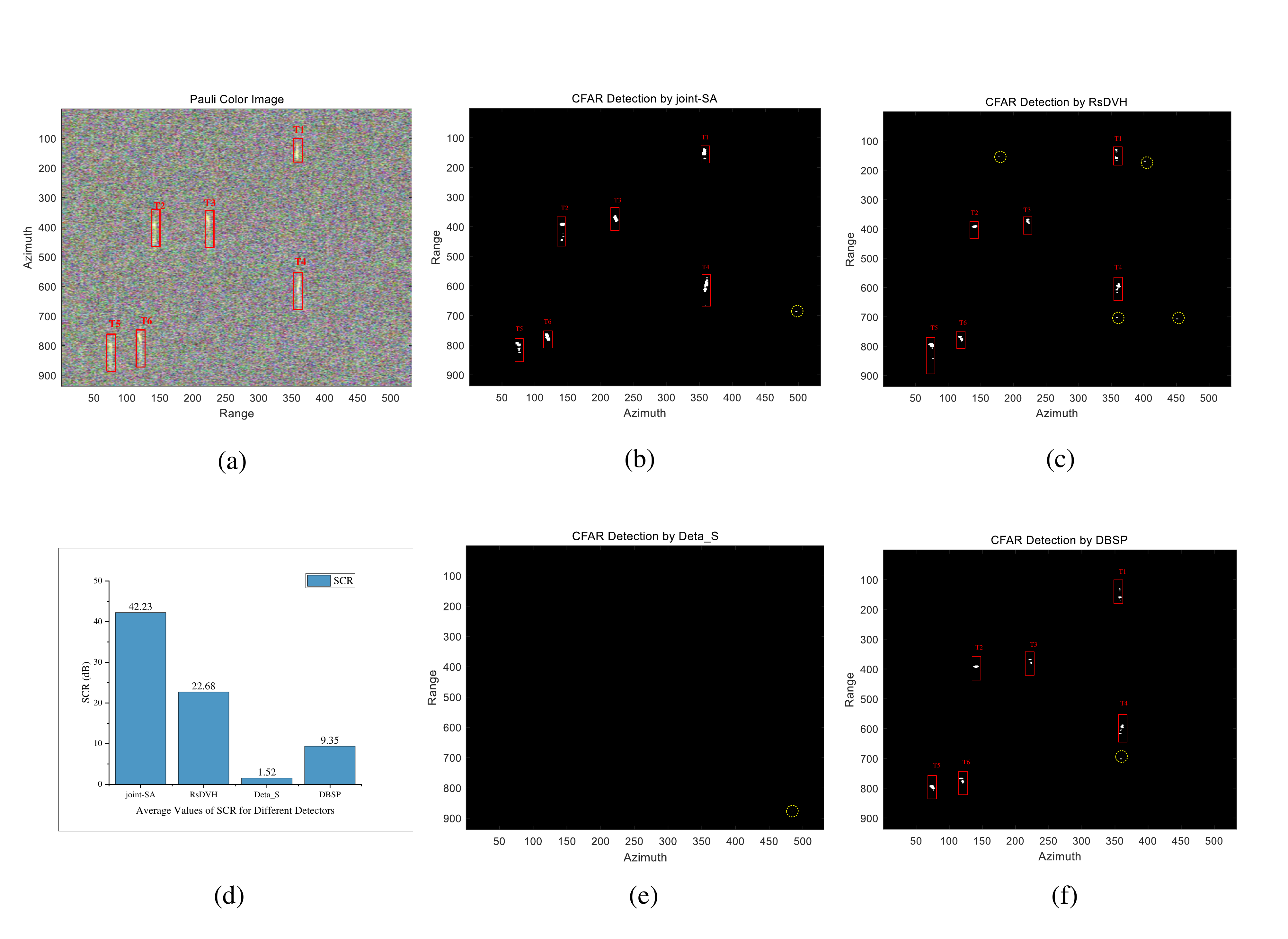}
	\caption{ Detection results of region C in scene 3. (a) Pauli RGB image. (b) CFAR detection result by joint-SA. (c) CFAR detection result by RsDVH. (d) Average values of SCR for different detectors. (e) CFAR detection result by $\Delta$S. (f) CFAR detection result by DBSP. Yellow dashed circles: false-alarm targets, and the red boxes: targets.}
	\label{fig:det_Region_C}
\end{figure}
\begin{figure}[H]
	\centering
	\includegraphics[width=\textwidth]{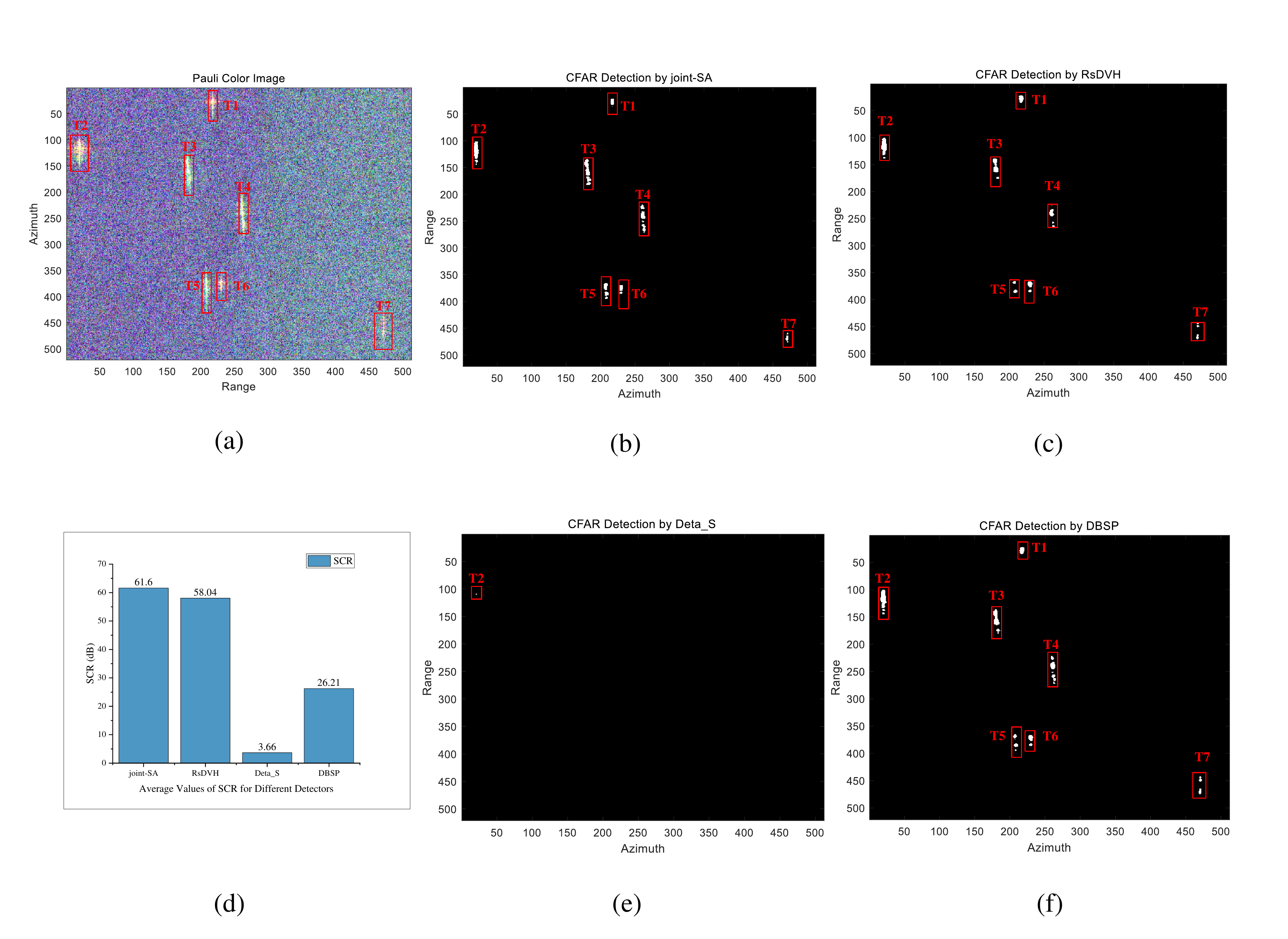}
	\caption{ Detection results of region E in scene 5. (a) Pauli RGB image. (b) CFAR detection result by joint-SA. (c) CFAR detection result by RsDVH. (d) Average values of SCR for different detectors. (e) CFAR detection result by $\Delta$S. (f) CFAR detection result by DBSP. Yellow dashed circles: false-alarm targets, and the red boxes: targets.}
	\label{fig:det_Region_E}
\end{figure}

To save space, we don’t analyze for region B in scene 2 and region E in scene 5 in detail and only show their detection result, as figures \ref{fig:det_Region_C}, \ref{fig:det_Region_E}, and tables \ref{tab:C_SCR}, \ref{tab:E_SCR}. Noted $\Delta$S that cannot effectively perform adaptive ship detection in these 5 scenes, so we do not analyze it. In summary, the proposed detector joint-SA is better than the other ship detectors (DBSP, RsDVH,  $\Delta$S, and so on) in terms of both the accuracy of ship detection and the SCR value, and more importantly, the detection performance of joint-SA does not vary with the incidence angles.

\begin{table}[h]
	
	\centering
	\caption{The SCR values for relatively weakly scattering targets in region C}
	\label{tab:C_SCR}

	\begin{tabular}{c|l|ccccccc}
		\hline
		\multicolumn{2}{c|}{Detectors}    & T1   &T2  &T3  &T4  & T5 &T6 \\ \hline
		\multicolumn{2}{c|}{RsDVH}    & 35.44   &54.28 &36.18  &39.25  &54.88 &42.37  \\ \hline
		\multicolumn{2}{c|}{$\Delta$S}    &0.95  &0.98  &0.81  &1.25  &1.45 &1.38 \\ \hline
		\multicolumn{2}{c|}{DBSP}    &14.43  &24.78  &15.02  &15.99  &24.88 &19.82  \\ \hline
		\multicolumn{2}{c|}{\textbf{joint-SA}}    &\textbf{36.42}  &\textbf{62.55}  &\textbf{44.82}  &\textbf{50.54}  &\textbf{61.70} &\textbf{50.19} \\ \hline
	\end{tabular}
\end{table}

\begin{table}[h]
	
	\centering
	\caption{The SCR values for relatively weakly scattering targets in region E}
	\label{tab:E_SCR}

		\begin{tabular}{c|l|ccccccc}
			\hline
			\multicolumn{2}{c|}{Detectors}    & T1   &T2  &T3  &T4  & T5 &T6 &T7\\ \hline
			\multicolumn{2}{c|}{RsDVH}    &57.92   &84.79  &52.95  &50.08  &47.52 &62.64 &28.48 \\ \hline
			\multicolumn{2}{c|}{$\Delta$S} &4.23  &2.02  &3.54  &4.04  &4.28  &4.54 &3.01 \\ \hline
			\multicolumn{2}{c|}{DBSP}    & 26.99  &37.59  &26.29  &24.58  &22.09 &28.27 &17.67 \\ \hline
			\multicolumn{2}{c|}{\textbf{joint-SA}}    &\textbf{64.26}  &\textbf{73.79}  &\textbf{62.82}  &\textbf{72.65}  &\textbf{59.24} &\textbf{65.31} &\textbf{33.16} \\ \hline
	\end{tabular}
\end{table}

\section{Conclusion and Discussion}
\label{s:Conclusion}
To realize the detection of relatively weakly scattering targets and improve the detection accuracy, we proposed a new full-polarization SAR image ship detector joint-SA, by joint the target scattering characteristics and the wave polarization anisotropy. First, we tested and verified the suitability of the G$\Gamma$D for characterizing joint-SA statistics of sea clutter with a wide range of homogeneity. Second, from the perspective of the SCR, we strictly demonstrated that the detector is an effective physical quantity to distinguish the ship from the sea clutter. Third, using the GF-3 in C-band type, the RADARSAT-2 in C-band type data, and the L-band ALOS data, we verified the correctness of the theory of the detector joint-SA and the superiority of the detection accuracy. \par
Because joint-SA considers two different characteristics: scattering characteristics and wave polarization anisotropy, the detector joint-SA can effectively improve SCR values and the accuracy of ship detection. Experiments were conducted using SAR data from different satellite platforms, wavebands, and incidence angles, and the experimental results proved that the performance of the detector joint-SA is better than the other three detectors when the incidence angles are less than 45°, and its performance does not vary with the incidence angle. That is to say, the robustness of joint-SA is better than the other three detectors. As a result, we can conclude that detector joint-SA is a flexible full-polarization ship detector that outperforms the other three detectors.\par
Since we lack data for incidence angles higher than 45°, we cannot verify the validity of joint-SA on high incidence angle SAR data. In future work, we should deal with these defects with other efficient characteristics. As for the detection performance in high sea state and incidence angles, we will continue to validate it when more PolSAR data and AIS data are available.
\section*{Acknowledgments}
This work was supported in part by the National Natural Science Foundation of China under Project 41822105, in part by Key Research Plan of Hunan Province under Project 2019SK2173, and in part by the Fundamental Research Funds for the Central Universities under Projects 2682020ZT34 and 2682021CX071.
\bibliographystyle{model2-names}
\bibliography{joint-SA}

\end{document}